\documentclass[12pt,a4wide,fleqn]{article}

\usepackage[authoryear]{natbib}
\usepackage{graphicx,color,enumerate,amssymb}
\usepackage{amsmath}
\usepackage{amssymb}
\usepackage{enumerate}
\usepackage{amsthm}
 \usepackage{bbm}
 \usepackage{booktabs}
\usepackage{xcolor}
 \usepackage{afterpage}
 
\renewcommand{\le}{\leqslant}
\renewcommand{\leq}{\leqslant}
\renewcommand{\ge}{\geqslant}
\renewcommand{\geq}{\geqslant}

\newtheorem{thm}{Theorem}[section]

\newtheorem{remark}[thm]{Remark}

\newcommand{\R}{\mathbb{R}}
\newcommand{\E}{\mathsf{E}\,}

\newcommand{\Prob}{\mathsf{P}}

\newcommand{\ind}{\mathbbm{1}}

\renewcommand{\d}{\mathrm{d}}

\newcommand{\omu}{\overline{X}_n^{\operatorname{O}}(u)}
\newcommand{\ohk}{\overline{h}_k^{\operatorname{O}}}
\newcommand{\ohka}{\overline{h}_{k,\operatorname{abr}}^{\operatorname{O}}}

\newcommand{\oY}{Y^{\operatorname{O}}}
\newcommand{\oYka}{Y^{\operatorname{O}}_{n,k,\operatorname{abr}}}
\newcommand{\oYkgra}{Y^{\operatorname{O}}_{n,k,\operatorname{gra}}}
\newcommand{\oYklin}{Y^{\operatorname{O}}_{n,k,\operatorname{lin}}}
\newcommand{\ow}{w^{\operatorname{O}}}
\newcommand{\owlin}{w^{\operatorname{O}}_{\gamma,\operatorname{lin}}}
\newcommand{\oomegalin}{\omega^{\operatorname{O}}_{\gamma,\operatorname{lin}}}

\newcommand{\oWka}{W^{\operatorname{O}}_{n,k,\operatorname{abr}}}
\newcommand{\oT}{T^{\operatorname{O}}}
\newcommand{\oTa}{T^{\operatorname{O}}_{n,\gamma,\operatorname{abr}}}
\newcommand{\oTag}[1]{T^{\operatorname{O}}_{n,#1,\operatorname{abr}}}
\newcommand{\hoTa}{\widehat{T}^{\operatorname{O}}_{n,\gamma,\operatorname{abr}}}
\newcommand{\hoTag}[1]{\widehat{T}^{\operatorname{O}}_{n,#1,\operatorname{abr}}}
\newcommand{\hoTlinS}{\widehat{T}^{\operatorname{O}}_{n,\gamma,\operatorname{lin},\operatorname{S}}}
\newcommand{\hoTlinSg}[1]{\widehat{T}^{\operatorname{O}}_{n,#1,\operatorname{lin},\operatorname{S}}}
\newcommand{\hoTlinI}{\widehat{T}^{\operatorname{O}}_{n,\gamma,\operatorname{lin},\operatorname{I}}}
\newcommand{\hoTlinIg}[1]{\widehat{T}^{\operatorname{O}}_{n,#1,\operatorname{lin},\operatorname{I}}}

\newcommand{\owmu}{\widehat{\mu}^{\operatorname{O}}}
\newcommand{\oZ}{Z^{\operatorname{O}}}
\newcommand{\oZka}{Z^{\operatorname{O}}_{n,k,\gamma,\operatorname{abr}}}
\newcommand{\oZklinS}{Z^{\operatorname{O}}_{n,k,\gamma,\operatorname{lin},\operatorname{S}}}
\newcommand{\oZklinI}{Z^{\operatorname{O}}_{n,k,\gamma,\operatorname{lin},\operatorname{I}}}
\newcommand{\oZkag}[1]{Z^{\operatorname{O}}_{n,k,#1,\operatorname{abr}}}
\newcommand{\woT}{\widehat{T}^{\operatorname{O}}}
\newcommand{\oo}[1]{#1^{\operatorname{O}}}

\newcommand{\tpi}{\boldsymbol{\pi}}

\begin{document}
\begin{center}
{ \LARGE\sc Detection of  mean changes in partially observed functional data \\
} 
\vspace*{0.5cm}  \large\sc  \v{S}\'arka Hudecov\'a$^a$ and Claudia Kirch$^b$\vspace*{0.5cm}
 
\today
\end{center}

\noindent$^a$ Department of Probability and Mathematical Statistics, Faculty of Mathematics and Physics, Charles University, Prague, Czech Republic,\\
email \verb"hudecova@karlin.mff.cuni.cz", ORCID  0000-0001-8135-7404,\\[1ex]
\noindent $^b$ Institute of Mathematical Stochastics (IMST), Department of Mathematics, Otto-von-Guericke University, Magdeburg, Germany,\\
email \verb"claudia.kirch@ovgu.de", ORCID 0000-0001-8135-7404

\vspace{0.5cm}

\paragraph{Abstract.}
We propose a test for a change in the mean for a sequence of functional observations that are only partially observed on subsets of the domain, with no information available on  the complement. The framework accommodates important scenarios, including both abrupt and gradual changes. The significance of the test statistic is assessed via a permutation test. In addition to the classical permutation approach with a fixed number of permutation samples, we also discuss  a variant with controlled resampling risk that relies on a random (data-driven) number of permutation samples. The small sample performance of the proposed methodology is illustrated in a Monte Carlo simulation study and an application to real data.

\section{Introduction}

Functional data analysis provides a rich framework for studying complex high-dimen\-sional stochastic observations that are regarded as infinite-dimensional objects (functions) rather than finite-dimensional vectors. In many applications, functional data are observed over time, such as  intraday volatility curves in finance \citep{KOKOSZKA2024104426}, or some environmental quantity such as daily (or yearly) curves of some pollutant levels, see \cite{KING2018233,hoermanndynamic15,Aue02012015} for examples based on particulate matter concentrations. In such examples, the functional nature of the observation captures the seasonal behavior of the data, thus avoiding a complex modeling of periodicity and non-stationarity. Other examples of functional data over a time course include neuroscience data sets in particular fMRI data \citep{aston2011,STOEHR202144}.

In all of these examples, a key question for reliable statistical inference and decision-making is whether the underlying data-generating process remains stable or undergoes changes, which may be either abrupt or gradual. Consequently, the problem of change point testing for functional data has attracted considerable attention over the past two decades, with applications ranging from environmental monitoring and biomedical studies to economics and finance. When considering a possible change in the mean, most of the works have focused on an abrupt change \citep{aue2009,aston2011,aston2012,bucchia2017, dette2020, aue2018}, while testing a gradual change has received less attention \citep{vogt,bucher,bastian}.

\begin{figure}[bp]
\centering
\includegraphics[width=0.7\textwidth]{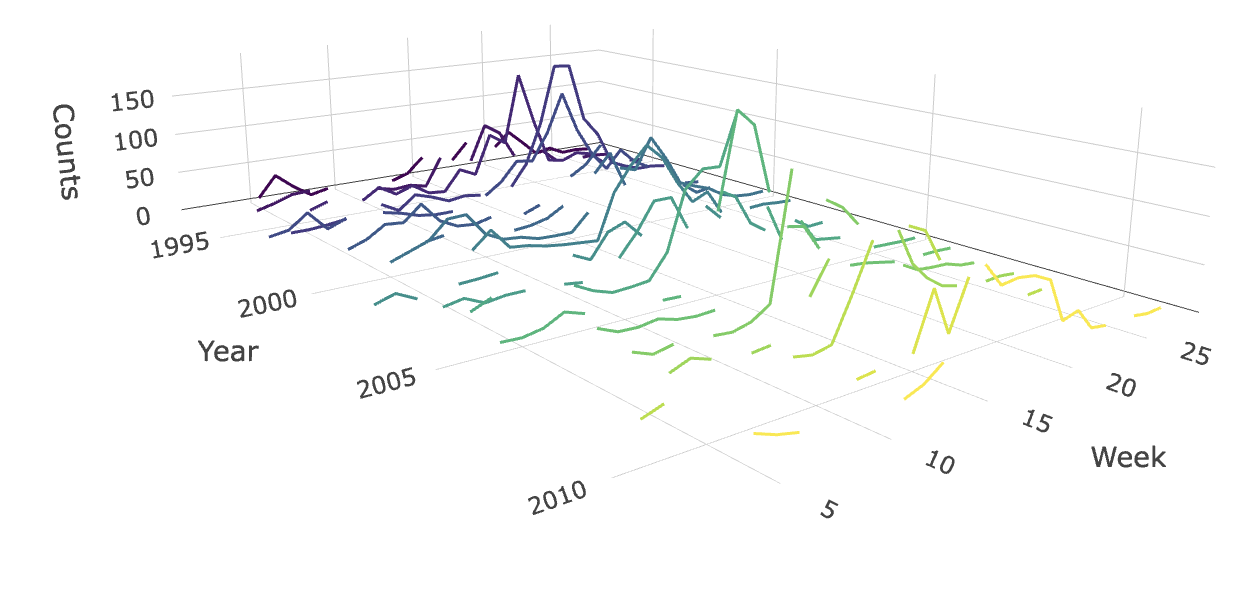}
\caption{Weekly butterfly counts at Drayton site, UK, in 1993--2011 from \cite{butterfly}. If the weather conditions did not meet the protocol, the measurement for that week is missing, so the number of missing observations within a year ranges between 3 and 18.  }\label{fig:data1}
\end{figure}

In some applications,  each function is observed only partially 
on a subset of the domain, with no information available on its values over the complement. For instance, in environmental monitoring, sensor failures or maintenance can lead to missing intervals in temperature or pollution curves. In medical studies, measured profiles may be incomplete if the patient removes the measurement device or if the device malfunctions \citep{kraus2015}.
In biology, certain measurements, such as animal counts, are recorded or considered reliable only under specific weather conditions. As an illustration,
Figure~\ref{fig:data1} 
presents weekly butterfly counts in Drayton, UK, from \cite{butterfly} over 19 consecutive years; see Section~\ref{sec:data} for a detailed description of this dataset. Each year, recordings were conducted weekly from early April to the end of September, provided that weather conditions satisfied the requirements of the measurement protocol. It is visible that the missingness in this dataset is substantial.

In this contribution, we consider the problem of testing for a mean change in functional data that are only partially observed. Standard change-point detection methods, which assume fully observed functional data, cannot be directly applied in such settings.
This motivates the development of statistical procedures that remain valid also for partially observed data. Our framework encompasses important special cases, including both abrupt and gradual changes. Since the length of the data sequence is often limited, as in the dataset in Figure~\ref{fig:data1}, we propose assessing the significance of the change point test statistic via a permutation test, which maintains its nominal level even in small samples under exchangeability under the null hypothesis. Because the theoretic permutation test is not accessible in practice, we discuss two implementations of this approach: the standard approximation of the conditional permutation distribution based on a fixed number of permutation samples, and a novel permutation approach, introduced by \cite{gandy2009} and later refined by \cite{gandy2020}. The latter approach continues to take permutation samples until one can be sufficiently certain (up to a pre-specified tolerance) that the test decision based on these permutation samples  equals the theoretic one, that is based on the true conditional permutation distribution. In effect, this procedure controls the resampling risk, i.e.\ the risk of reaching a different decision with the approximation as compared to the theoretic one. By construction, the number of permutation samples is random depending on the data at hand but also the permutation samples taken, where typically much fewer samples are required as compared to the classical approach.

The paper is structured as follows: The model and the class of test statistics are introduced in Section~\ref{sec:2}. Section~\ref{sec:3} focuses on the important special cases of an abrupt and a gradual change. Practical aspects, including the two variants of permutation tests, are discussed in Section~\ref{sec:4}. The small sample performance of the proposed methodology is illustrated by a Monte Carlo simulation study in Section~\ref{sec:simul}. The analysis of the butterfly counts from  Figure~\ref{fig:data1} is provided in Section~\ref{sec:data}.

\section{Mean change in partially observed functional data}\label{sec:2}

Let $L^2(D)$ be the set of all square-integrable functions defined on a compact set $D\subset \R^d$ for some $d\geq 1$. Then,  $L^2(D)$  is a separable Hilbert space equipped with the inner product $\langle f,g \rangle = \int_{D} f(u)g(u)\d u$ for $f,g \in L^2(D)$ and the corresponding norm $\|f\| = \sqrt{\int_{D} f^2(u)\d u}$. 
A functional random variable $X$ is defined on a probability space $\Omega$ and takes values in  $L^2(D)$. 

In practical applications, one often deals with a sequence of functions $X_1,\dots,X_n$ and the aim is to decide whether the data generating mechanism is stable over time. 
We consider the setup, where $X_i$ is only partially observed on a set $\mathcal{O}_i\subset D$, and no information about $X_i(u)$ is available for $u\not\in \mathcal{O}_i$.
 For $u\in D$,
set $O_i(u) = \ind_{\mathcal{O}_i}(u)$, the indicator that $X_i(u)$ is observed. 
The available data consist of 
\[
\mathcal{X} = \Big\{ \{ X_i(u), \  u\in \mathcal{O}_i\}, \quad i=1,\dots,n \Big\}.
\]
In what follows, assume that the observation sets $\mathcal{O}_1,\dots,\mathcal{O}_n$ are mutually independent, identically distributed (i.i.d.)
and independent of $X_1,\dots,X_n$.


\subsection{Model}

Let $\eta_1,\dots,\eta_n$ be i.i.d. functional random variables such that 
$\E \eta_i = 0$ and assume that $X_i$ satisfies
\begin{equation}\label{model:general}
X_i(u) = \mu(u) + \delta(u)\cdot g_u(i/n) + \eta_i(u), \quad i=1,\dots,n,
\end{equation}
for $u\in D$, where $\mu:D\to\R$ and $\delta:D\to\R$ are square integrable functions  and  $g_u:[0,1]\to\R^+$ is a function, in the following referred to as \emph{change function},  that describes how the 
mean of the observations is evolving over time at location $u$.  In this work, we will 
use monotonically increasing change functions that do not depend on the location $u$, i.e.\ the mean evolves in the same manner at all locations except for the \emph{magnitude} $\delta(u)$.
We are interested in testing the null hypothesis of no change 
\[
H_0: \E X_1 = \dots = \E X_n
\]
where the equality holds for almost all $u\in D$ with respect to the Lebesgue measure on $D$, 
against a general alternative.

In this work, we will concentrate on test statistics defined for the situation of at most one change point with the same structure of the change for all point in $u$, i.e.\ where 
$g_u(t)=h(t-\kappa)$ for some monotonically increasing function $h:[-1,1]\to\R$ with $h(x)=0$ for $x\le 0$. Here, $\kappa \in [0,1]$ is the unknown change point in rescaled time. The fact that $g$ is constant throughout $u$ reflects the situation that we expect the same time of change as well as the same dynamic (e.g., abrupt, gradual with linear slope, gradual with quadratic slope etc.) across all points $u$. Nevertheless, this allows for some points to not exhibit a change at all (via $\delta(u)=0$), some means to increase ($\delta(u)>0$) and other to decrease ($\delta(u)<0)$ with time. However, the setup does not allow for a gradual linear changes in some region and an abrupt change in another.  In situations, where previous knowledge about the anticipated type of change in different regions is available, the proposed test procedures can easily be generalized to that context.
Yet, in general, such previous knowledge cannot be expected to be available. Indeed, even knowledge about the true shape of $g_u$ can usually not be expected, which is why we will also investigate the power of the test for misspecified situations, where the test statistic is designed based on a given fixed $h$, but the power is not only investigated in the situation where this is the true shape of the change function but also for situations where $g_u(\cdot)\neq h(\cdot-\kappa)$ for all $0<\kappa<1$, where $g_u$ denotes the true change function, see Section~\ref{sec:simul:G}.

While we concentrate on the at-most-one-change situation, both an abrupt and a gradual one, in this paper, extensions to more complex situations are also straightforward, see e.g.\ Remark~\ref{rem_epidemic} for an explanation on how this is done for so called epidemic changes.

\subsection{Test statistic}

The traditional test statistic  for fully observed data designed for an at-most-one-change (AMOC) of shape $h$ is given by  
\begin{align}\label{eq_stat_general}
T_{n,h,\nu_{\gamma,h}}=\max_{1\leq k <n} \nu_{\gamma,h} 
\left(k,n\right) 
\cdot \|Y_{n,k,h}\|^2=\max_{1\le k<n} \|Z_{n,k,h,\nu_{\gamma,h}}\|^2,
\end{align}
where $Z_{n,k,h,\nu_{\gamma,h}}=\sqrt{\nu_{\gamma,h} 
\left(k,n\right) }\, Y_{n,k,h}$ and
\begin{align}\label{eq_general_cusum}
Y_{n,k,h}(u) = \frac{1}{\sqrt{n}}\sum_{i=1}^n h\left(\frac{i-k}{n}\right) \left(X_i(u) - \overline{X}_n(u)\right), \quad u\in D,
\end{align}
and $\nu_{\gamma,h}$ is some non-negative weight function  depending on a parameter $0<\gamma\le 1/2$ and the function $h$. We will discuss the choice of the weight function in more detail in Section~\ref{sec:w}.
The null hypothesis is rejected for large values of the test statistic. 

The test statistic from \eqref{eq_stat_general} cannot be directly used for partially observed functional data, because the information about $\{X_i(u), u\not\in \mathcal{O}_i\}$ is missing, so a careful modification of \eqref{eq_stat_general} is necessary.

 For $u\in D$,
introduce the notation 
\[
N(u) = \sum_{i=1}^n O_i(u), \quad 
N_k(u) =\sum_{i=1}^k O_i (u),  \qquad 1\leq k \leq n.
\]
We suggest to replace the functional variable $\{Y_{n,k,h}(u), u\in D\}$ by 
$\{\oY_{n,k,h}(u), u\in D\}$ where, with the convention $\frac 0 0=0,$
\begin{align}\label{eq_cusum_general_missing1}
\oY_{n,k,h}(u) &= \frac{ \ind[N(u)>0]}{\sqrt{N(u)}}\sum_{i=1}^n h\left(\frac{i-k}{n}\right) O_i(u) \left(X_i(u) -\omu\right)\\
\label{eq_cusum_general_missing2}
&= \frac{ \ind[N(u)>0]}{\sqrt{N(u)}}\sum_{i=1}^n h\left(\frac{i-k}{n}\right) O_i(u) X_i(u) - \sqrt{N(u)}\cdot \ohk (u) \cdot \omu, 
\end{align}
where
\begin{align*}
\omu &= \ind[N(u)>0] \cdot \frac{1}{N(u)}\sum_{i=1}^n O_i(u) X_i(u),\\
\ohk(u) &= \ind[N(u)>0] \cdot \frac{1}{N(u)} \sum_{i=1}^n h\left(\frac{i-k}{n}\right) O_i(u).  
\end{align*}
Note that $\omu$ is the estimator of the mean of a partially observed random sample proposed by \cite{kraus2015}. 
Subsequently, we propose the test statistic 
\begin{equation}\label{eq_stat_missing}
\oT_{n,h,\ow_{\gamma,h}}=\max_{1\leq k <n}  \|\oZ_{n,k,h,\ow_{\gamma,h}}\|^2 =  \max_{1\leqslant k <n} \int_{D}  \left(\oZ_{n,k,h,\ow_{\gamma,h}}(u)\right)^2 \mathrm{d} u,
\end{equation}
where $\oZ_{n,k,h,\ow_{\gamma,h}}(u)$ is the weighted version of $\oY_{n,k,h}(u)$ given by
\begin{equation}\label{eq:Z:gen}
\oZ_{n,k,h,\ow_{\gamma,h}}(u) = 
\sqrt{\ow_{\gamma,h}(k,n;u)}
\,\oY_{n,k,h}(u) \cdot \ind[0<N_k(u)<N(u)]
\end{equation}
 for some non-negative weight function $\ow_{\gamma,h}$. 
 In contrast to $\nu_{\gamma,h}$ the function $\ow_{\gamma,h}$ can depend on the missingness, so it is generally a function of $u\in D$ and, therefore, it cannot be taken out of the norm as in \eqref{eq_stat_general} resulting also in a higher computational burden. 
 The choice of the weight function is discussed in Section~\ref{sec:w}. 

The null hypothesis $H_0$ is rejected for large values of $\oo{T}_{n,h,\ow_{\gamma,h}}$, and we recommend to assess the significance of the test statistic 
using a permutation test with bounded resampling risk, an approach described in detail in Section~\ref{sec:perm2}.


As the test statistic in \eqref{eq_stat_missing} is designed with an at-most-one-change (AMOC)  of shape $h$ in mind, it can be expected to have best power behavior if the true $g_u$ is close in shape to $h(\cdot-\kappa)$ for some $\kappa=k/n$. However, it also has power in other situations, as investigated in more detail in Section~\ref{sec:simul:G}.

\subsection{Weight function}\label{sec:w}

 For fully observed univariate data, \cite{huskova2000} consider the test statistic in \eqref{eq_stat_general} together with the weight function $\nu_{1/2,h}$, where
\begin{equation}\label{eq:nu}
\nu_{\gamma,h}(k,n) =  \left\{\frac{1}{n}\sum_{i=1}^n h\left(\frac{i-k}{n}\right)^2 - \left[ \frac{1}{n} \sum_{i=1}^n   h\left(\frac{i-k}{n}\right)\right]^2 \right\}^{-2\gamma}, \qquad 0\le \gamma\le \frac 1 2,
\end{equation}
provides a generalization of their approach. 
This motivates us to consider the weight function $\ow_{\gamma,h}$ for the test statistic \eqref{eq_stat_missing} in the form  
\begin{align}\label{eq:w-1}
\ow_{\gamma,h}(k,n;u) &= \left\{\frac{1}{N(u)} \sum_{i=1}^n h\left(\frac{i-k}{n}\right)^2 O_i(u)- \left[\frac{1}{N(u)} \sum_{i=1}^n  h\left(\frac{i-k}{n}\right)O_i(u)\right]^2 \right\}^{-2\gamma}
\end{align}
for $\gamma\in[0,1/2]$. 
An integral approximation gives $\nu_{\gamma,h}(k,n) \approx \omega_{\gamma,h}(k/n)$  for large $n$ with
\begin{equation}\label{eq_weight_h}
\omega_{\gamma,h}(t)=\left\{\int_0^{1-t} h^2(z) \mathrm{d} z - \left[\int_0^{1-t} h(z) \mathrm{d} z \right]^2\right\}^{-2\gamma},\qquad 0\leqslant \gamma\le 1/2.
\end{equation}
Therefore, another possibility is to use the test statistic from \eqref{eq_stat_missing} with an approximate weighting
\begin{equation}\label{eq:w-2}
\ow_{\gamma,h}(k,n;u) = \omega_{\gamma,h}\left(\frac{k}{n}\right),
\end{equation}
where $\omega_{\gamma,h}$ is as in \eqref{eq_weight_h}. An advantage of this simplified approach is that the weights from \eqref{eq:w-2} do not depend  on $u$, such that the weight function can be taken out of the integral as in \eqref{eq_stat_general}, providing a global rather than location- and data-dependent weight. This is particularly important from a computational perspective, especially when the significance of $\oT_{n,\ow_{\gamma,h}}$ is assessed using resampling methods, see also the discussion in Section~\ref{sec:simul}.




\begin{remark}
For fully observed data, where $\mathcal{O}_i=D$ for all $i=1,\dots,n$, we have $N(u)=n$, $N_k(u)=k$ and $\oY_{n,k,h}(u)=Y_{n,k,h}(u)$ for all $u\in D$. Consequently, $\oT_{n,\ow_{\gamma,h}}$ with 
$\ow_{\gamma,h}$ from \eqref{eq:w-1} yields
 the same  test statistic as $T_{n,\nu_{\gamma,h}}$ in \eqref{eq_stat_general} with weights $\nu_{\gamma,h}$ from~\eqref{eq:nu}. Consequently, the standard inference for fully observed functional data is included  in the current framework.
\end{remark}


\section{Important special cases}\label{sec:3}

Two important special cases of model \eqref{model:general} are discussed below: Section~\ref{sec:amoc} discusses the statistic constructed with an abrupt change in mind, while Section~\ref{sec:gradual} discusses the statistic corresponding to a gradual change.

\subsection{An abrupt change}\label{sec:amoc}


An abrupt change occurs when the data-generating process undergoes a sudden shift from one state to another. The classical approach to this problem, based on cumulative sums, was introduced by \citet{page}, and has since been extended in various directions to accommodate more complex frameworks \citep{CH,chen-gupta}. For a recent overview on abrupt-change point detection in fully observed functional data see \citet[Chapter 8]{horvath2024}.

\begin{figure}[tbp]
 \centering
 \includegraphics[width=0.47\textwidth]{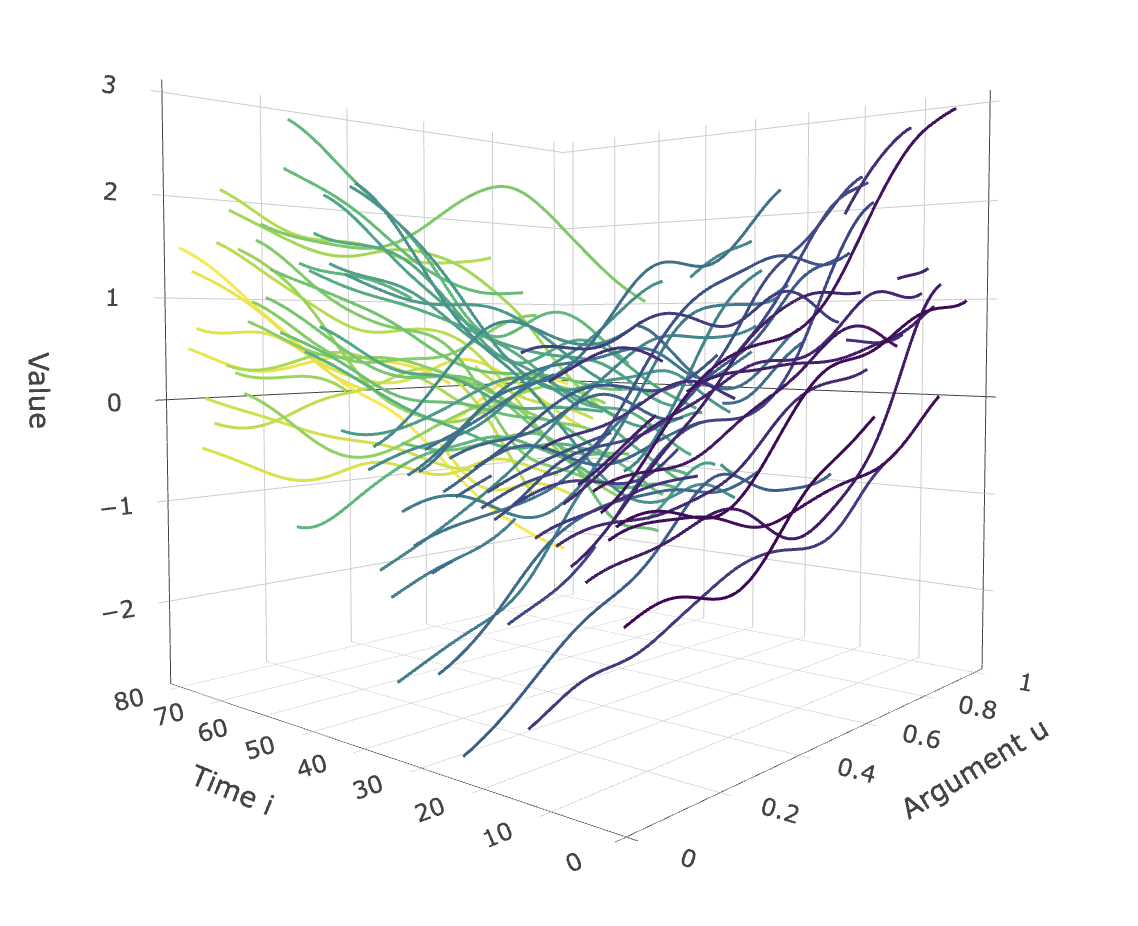}
  \includegraphics[width=0.45\textwidth]{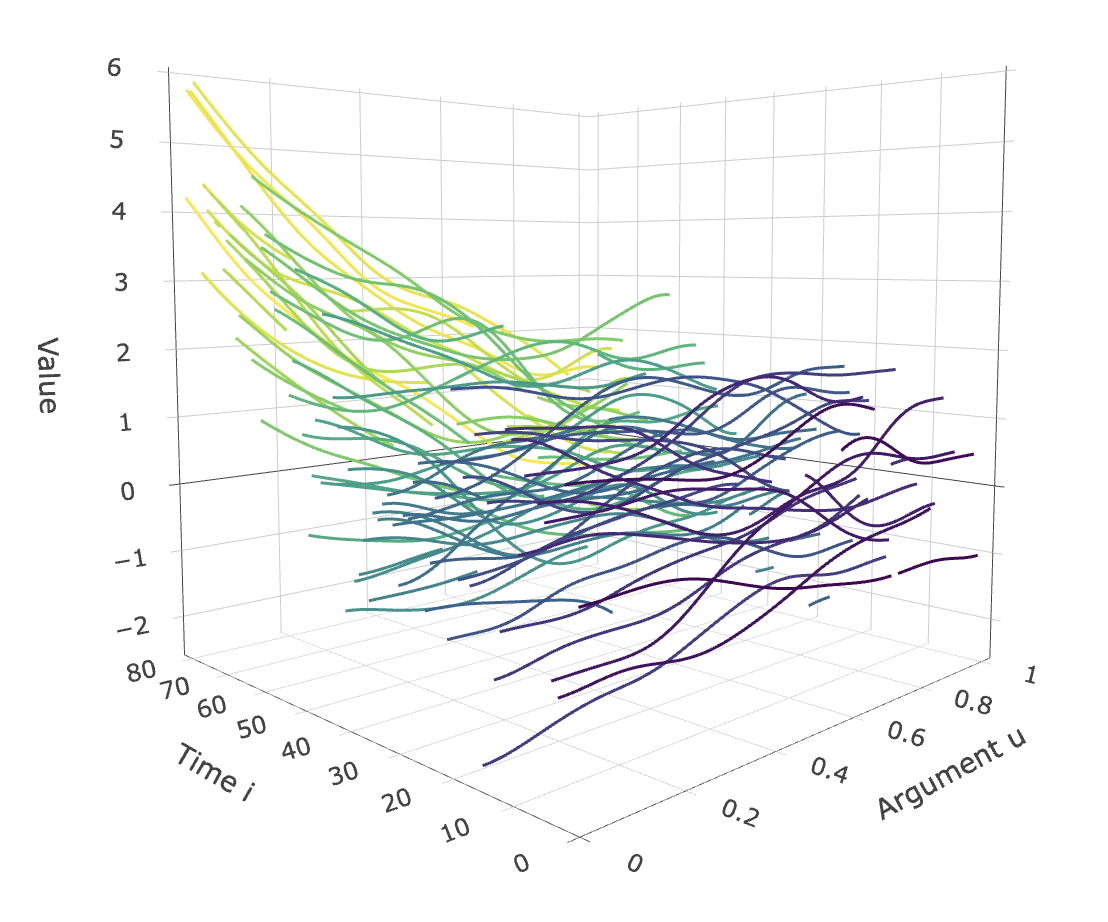}
 \caption{An example of a partially observed data $X_1,\dots,X_n$ for $n=80$ and $D=[0,1]$ with an abrupt change at $k=n/2$ (left plot) and with a linear gradual change with $\kappa=1/2$ (right plot).}\label{fig:example1}
 \end{figure}

In the context of the general model \eqref{model:general}, the AMOC abrupt change is obtained for 
 $g_u(t) = h_{\operatorname{abr}}(t-\kappa)$  for some $0<\kappa<1$ with $h_{\operatorname{abr}}(x) = \ind[x>0]$.
In that case, for $u\in D$,
\[
\E X_i(u) = \mu(u), \ 1\leqslant i\leqslant k, \qquad \E X_i(u) = \mu(u)+\delta(u),\quad k<i\leqslant n,
\]
where $k = \lfloor n \kappa \rfloor$ is the unknown change point. The left panel of Figure~\ref{fig:example1} provides an illustration of a corresponding functional partially observed dataset $\mathcal{X}$ with $\|\delta\|>0$.

Applying the function $h_{\operatorname{abr}}(x)$ to the general formula \eqref{eq_cusum_general_missing2}   leads to 
\[
\ohka (u)= \frac{1}{N} \sum_{i=k+1}^n O_i(u) = \frac{N(u)-N_k(u)}{N(u)}, 
\]
and  $\oY_{n,k,h}$ from \eqref{eq_cusum_general_missing1} takes form
\begin{align*}
\oYka(u)
& = \frac{\ind[N(u)>0]}{\sqrt{N(u)}} \sum_{i=k+1}^n O_i(u)X_i(u) - \sqrt{N(u)} \frac{N(u)-N_k(u)}{N(u)} \omu. 
\end{align*}
Since
\[
\omu = \frac{1}{N(u)} \left[ N_k(u) \,\owmu_{1,k}(u) + (N(u)-N_k(u))\, \owmu_{2,k}(u) \right],
\]
where
\begin{align*}
\owmu_{1,k}(u)& =  \frac{\ind[N_k(u)>0]}{N_k(u)}\sum_{i=1}^k O_i(u) X_i(u),\notag\\
\owmu_{2,k}(u)& =  \frac{\ind[N(u)-N_k(u)>0]}{N(u) - N_k(u)}\sum_{i=k+1}^n O_i(u) X_i(u), 
\end{align*}
we get the equality $\oYka(u)=-\oWka(u)$, where
\begin{align}\label{eq_W_amoc} 
\oWka(u)
 &= 
 \sqrt{N(u)} \left(\frac{N_k(u) [N(u)-N_k(u)]}{N^2(u)}\right)  \cdot \left[ \owmu_{1,k}(u) - \owmu_{2,k}(u) \right],
\end{align}
 with the convention that $\frac 00=0$.
Hence,  $\oWka$ (and so $\oYka$) can  be expressed as a difference of the estimators for the mean of partially observed functional data, 
constructed from $\{X_i,\ i\leqslant k\}$ and $\{X_i,\ i>k\}$ respectively.
The weight function from \eqref{eq:w-1} reduces to 
\[
\ow_{\gamma,\operatorname{abr}}(k,n;u)= w_{\gamma,\operatorname{abr}}(N_k(u)/N(u))
\]
where
\begin{align}\label{eq_weight}
w_{\gamma,\operatorname{abr}}(t) &= [t(1-t)]^{-2\,\gamma}\cdot \ind[0<t<1],\qquad 0\leqslant \gamma\le 1/2,
\end{align}
which is a standard weight function considered for an abrupt change point detection, see \citet[Chapter 1]{horvath2024}. Here,  the approximate weighting from \eqref{eq:w-2} yields 
$ w_{\gamma,\operatorname{abr}}(k/n)$. 

To sum up, $H_0$ is rejected for large values of $\oTa$, which equals the statistic in \eqref{eq_stat_missing} based on 
\begin{equation}\label{eq:Z:abr}
\oZka(u) 
= 
\sqrt{N(u)} \left(\frac{N_k(u) [N(u)-N_k(u)]}{N^2(u)}\right)^{1-\gamma}  \cdot \left[ \owmu_{1,k}(u) - \owmu_{2,k}(u) \right].
\end{equation}
The locations, where the statistics attain their maximum, seem to be good estimators for the time of the change if the change function is correctly specified, see Section~\ref{sec:estimator} in the Supplementary file.

\begin{remark}
For fully observed data, where $\mathcal{O}_i=[0,1]$ for all $i=1,\dots,n$, we get
$\oWka(u) = W_n(k/n,u)$, 
where
\begin{align} 
W_n(t,u) &= \frac{1}{\sqrt{n}}\left(\sum_{i=1}^{\lfloor nt \rfloor}X_i (u) - \frac{\lfloor nt\rfloor }{n}\sum_{i=1}^n X_i(u) \right) \notag \\ 
& = \sqrt{n} \frac{\lfloor nt \rfloor (n-\lfloor nt \rfloor)}{n^2} \left(\frac{1}{\lfloor nt \rfloor} \sum_{i=1}^{\lfloor nt \rfloor}X_i (u) - \frac{1}{n-\lfloor nt \rfloor}\sum_{i=\lfloor nt \rfloor +1 }^{n}X_i (u)\right) \label{eq:Wn}
\end{align}
is the classical functional CUSUM process, see \citet[Chapter 8.1]{horvath2024}.
\end{remark}

\begin{remark}
\cite{kraus2019} proposed a test for equality of means in $K$ samples of partially observed data with $D=[0,1]$. If we take  $\{X_i,\ i\leqslant k\}$ and $\{X_i,\ i>k\}$ as two independent samples and apply this test for $K=2$ (with common covariance operator), then the corresponding test statistic of \cite{kraus2019} is proportional to
\begin{align}
\Big\| \sqrt{N_k(\cdot)}\big[\owmu_{1,k}(\cdot)&- \overline{X}^{\operatorname{O}}_n(\cdot)\big]\Big\|^2 +  \Big\| \sqrt{N(\cdot) - N_k(\cdot)}\big[\owmu_{2,k}(\cdot)- \overline{X}^{\operatorname{O}}_n(\cdot)\big]\Big\|^2 \label{eq:kraus}.
\end{align}
Since $\owmu_{1,k}(u)- \overline{X}^{\operatorname{O}}_n(u)= [N(u)-N_k(u)]/N(u) \cdot [\owmu_{1,k}(u) - \owmu_{2,k}(u)]$ and $\owmu_{2,k}(u)- \overline{X}^{\operatorname{O}}_n(u)= N_k(u)/N(u) \cdot [\owmu_{2,k}(u) - \owmu_{1,k}(u)]$, \eqref{eq:kraus} can be rewritten as
\[
\left\|\frac{N(u)}{\sqrt{N_k(u)[N(u)-N_k(u)]}}\oWka(u)\right\|^2 = \|\oZkag{1/2}\|^2
\]
where $\oWka(u)$ is in \eqref{eq_W_amoc} and $\oZkag{1/2}(u)$ is its weighted version from \eqref{eq:Z:abr} obtained for $\gamma=1/2$. Hence, the change point test statistic $\oTag{1/2}$
is given by  the maximum over these two-sample test statistics
for all $k=1,\dots,n-1$. 
\end{remark}

\begin{remark}\label{rem_epidemic}
Let us shortly explain, how generalizations beyond the at-most-one-change situation can be obtained, using the example of an abrupt epidemic change for which $g_u(t)=h(t;\kappa_1,\kappa_2)=\ind[\kappa_1<t\le \kappa_2]$ for two unknown change points $\kappa_j$, $j=1,2$ (in rescaled time). In this setting, the mean function equals $\mu(u)$ except for $t\in (\kappa_1,\kappa_2]$, for which the mean function changes to $\mu(u)+\delta(u)$. The standard statistic for fully observed data based on an epidemic change  is based on a weighted maximum over $1\le k_1<k_2\le n$ of
\begin{align*}
    &\frac{1}{\sqrt{n}}\sum_{i=1}^{n}h\left(\frac{i}{n};\frac{k_1}{n};\frac{k_2}{n}\right)\,\left(X_i(u) - \overline{X}_n(u)\right)
    =\frac{1}{\sqrt{n}}\sum_{i=k_1+1}^{k_2}\left(X_i(u) - \overline{X}_n(u)\right).
\end{align*}
\cite{aston2012} discuss a version of such a statistic for fully observed functional data and dimension-reduction techniques.
Extensions to the partially observed situation can then be obtained analogously to what was outlined in this section.
\end{remark}

\subsection{A gradual change}\label{sec:gradual}

The setup considered in the previous section allows for a single abrupt change. 
However, in many real-world applications, the assumption of abrupt changes is overly restrictive, because the underlying processes often evolve gradually rather than through sudden jumps. In such cases, modeling the data-generating mechanism with a smoothly varying mean function may provide a more realistic description. %
This has led to a growing body of work on gradual changes, with important contributions including, e.g. \cite{huskova1999},  \cite{huskova2000}, \cite{vogt}, \cite{bucher}, \cite{bastian}.

A gradual at-most-one change model (with a change function homogeneous in $u$) is obtained in model \eqref{model:general} for   $g_u(t) = h_{\operatorname{gra}}(t-\kappa)$, where $h_{\operatorname{gra}}$ is a continuous function such that $h_{\operatorname{gra}}(x)=0$ for $x\leq 0$  and $h_{\operatorname{gra}}$ is increasing for $x>0$.
Here $k=\lfloor n\cdot \kappa \rfloor$ is the unknown change point. In this case, for $u\in D$,
$\E X_1(u)= \dots =\E X_k(u) = \mu(u)$ and 
\[
\E X_i(u) = \mu(u) + \delta(u) \cdot h_{\operatorname{gra}}\left(\frac{i-k}{n}\right), \quad \text{ for } i>k.
\]
Consequently, under alternatives, $\E X_i(u)$ is monotonous in $i$ for every $u$, where it may be increasing for some $u$ and decreasing or even constant for others. The right panel of Figure~\ref{fig:example1} presents  an example of the corresponding partially observed functional dataset with $\|\delta\|>0$.

The process 
in \eqref{eq_cusum_general_missing1},  designed with such a change function $h_{\operatorname{gra}}$ in mind, takes the form
\[
\oYkgra(u) = \frac{ \ind[N(u)>0]}{\sqrt{N(u)}}\sum_{i=k+1}^n h_{\operatorname{gra}}\left(\frac{i-k}{n}\right) O_i(u) \left(X_i(u) -\omu \right)
\]
for $u\in[0,1]$,  and the test statistic $\oT_{n,h_{\operatorname{gra}},w_{\gamma,h_{\operatorname{gra}}}}$ from \eqref{eq_stat_missing} can be combined either with the weighting in \eqref{eq:w-1} or the approximate weights in \eqref{eq:w-2}.


A standard class of these test statistics
is given by change functions $h_{\operatorname{pol},r}(x)=(x)_+^{r}$ for some $r> 0$, where $(x)_+ = x \cdot \ind[x>0]$ is the positive part of $x\in \R$. 
Formally, using $r=0$ leads back to the abrupt change situation in Section~\ref{sec:amoc}.
 For  $h_{\operatorname{pol},r}$ with $r>0$, the weight function in \eqref{eq:w-1} takes the form
\begin{equation*}
\left\{\frac{1}{N(u)} \sum_{i=k+1}^n  \left(\frac{i-k}{n}\right)^{2r} O_i(u)-   \left[\frac{1}{N(u)} \sum_{i=k+1}^n \left(\frac{i-k}{n}\right)^rO_i(u)\right]^2 \right\}^{-2\gamma}, 
 \end{equation*}
$0\leqslant \gamma\leqslant1/2$. In the simulations, we focus on 
tests designed for the special case for a linear change function $h_{\operatorname{lin}}(x)=h_{\operatorname{pol},1} =(x)_+$ with $r=1$, where \eqref{eq:w-1} takes the form
\begin{equation}
\label{eq:w-grad1}
\owlin(k,n;u) = 
\left\{\frac{1}{N(u)} \sum_{i=k+1}^n  \left(\frac{i-k}{n}\right)^{2} O_i(u) -   \left[\frac{1}{N(u)} \sum_{i=k+1}^n \left(\frac{i-k}{n}\right) O_i(u)\right]^2 \right\}^{-2\gamma}. 
 \end{equation}
The approximate weight function $\omega_{\gamma,h}$ from \eqref{eq_weight_h}
reduces to 
\begin{equation*}
\left\{(1-t)^{2r+1}\left[\frac{1}{2r+1} - \frac{1}{(r+1)^2}(1-t)\right]\right\}^{-2\gamma},
\end{equation*}
$0\leqslant \gamma\leqslant1/2$, with the special case for a linear change function  with $r=1$ being
\begin{equation}\label{eq_weight_lin}
\oomegalin\left(\frac k n \right)=
\left\{ \frac{(n-k)^3}{12 n^4} (3k+n)\right\}^{-2\gamma}. 
\end{equation}
%
Furthermore, for linear change functions, the process $\oY_{n,k,h}$ from \eqref{eq_cusum_general_missing1} takes the form
\begin{align*}
\oYklin(u) =& \frac{ \ind[N(u)>0]}{\sqrt{N(u)}}\left\{\frac{1}{n}\sum_{i=k+1}^n i\, O_i(u) X_i(u) -\frac{1}{n}\omu\sum_{i=k+1}^n i\, O_i(u) 
\right.\\
&\left.- \frac{k}{n} \sum_{i=k+1}^n O_i(u)X_i(u)+ \frac{k\cdot [N(u) - N_k(u)]}{n} \,\omu\right\}\notag
\end{align*}
and $\oZklinS=\oZ_{n,k,h_{\operatorname{lin}},\owlin}$, i.e.\ $\oZ_{n,k,h,\ow_{\gamma,h}}$ from \eqref{eq:Z:gen} is combined  with the \emph{sum-type} weights $\owlin$ from  \eqref{eq:w-grad1}. Alternatively, we can combine it with the \emph{integral-type} weights $\oomegalin$ as in \eqref{eq_weight_lin}, 
which we denote by $\oZklinI=\oZ_{n,k,h_{\operatorname{lin}},\oomegalin}$.

\section{Implementation of the test} \label{sec:4}

In this section, we discuss some aspects of the practical implementation of the above test procedures including the approximation of the integral and an efficient implementation of the permutation methodology.
\subsection{Practical computation of the test statistics}

In practice, functional data are represented via their discretized version.  
We assume that the data are recorded on a grid $\{u_j\}_{j=1}^q$ with  $u_j\in D\subset \R^d$. 
In particular, each function $X_i$ is measured solely at points $\{u_j\}_{j=1}^q \cap \mathcal{O}_i$, meaning that observations 
\[
\mathcal{X}_D = \big\{ \{X_i(u_j),\  u_j\in \mathcal{O}_i\big\}, \ i=1,\dots,n\}
\]
are available, where we assume without loss of generality that
$N(u_j)>0$ for all $j=1,\ldots,q$. 
Subsequently, the function $\oZ_{n,k,h,\ow_{\gamma,h}}$ as in \eqref{eq:Z:gen} is also available only at data points $\{u_j\}_{j=1}^q $.
The  integral in the test statistic $\oT_{n,h,\ow_{\gamma,h}}$ as in \eqref{eq_stat_missing} can be approximated using the Voronoi partition of $D$
 induced by the sample points $u_1,\dots,u_q$, see, e.g., \cite{lautensack}, as
 \[
\|\oZ_{n,k,h,\ow_{\gamma,h}}\|^2 \approx \sum_{j=1}^q \oZ_{n,k,h,\ow_{\gamma,h}}(u_j)^2 \cdot \mathrm{vol}(D\cap V_j), 
 \]
where $V_j$ is the Voronoi cell of $u_j$, $j=1,\dots,q$, and $\mathrm{vol}(A)$ denotes the Lebesgue measure of a set $A \subset \R^d$. 

For the simplest case $d=1$ and $D=[0,1]$,  this approximation takes the form
\begin{equation}\label{eq:aprox}
\|\oZ_{n,k,h,\ow_{\gamma,h}}\|^2 \approx \sum_{j=1}^{q}\oZ_{n,k,h,\ow_{\gamma,h}}(u_j)^2  (v_{j}-v_{j-1}),
\end{equation}
where $v_0=0, v_j=\frac{u_{j}+u_{j+1}}{2}$ for $j=1,\ldots,q-1$, and $v_q=1$. In this case, 
\begin{align*}
\sum_{j=1}^{q}&\oZ_{n,k,h,\ow_{\gamma,h}}(u_j)^2  (v_{j}-v_{j-1}) \\
&= \frac{1}{2} \sum_{j=1}^q  \oZ_{n,k,h,\ow_{\gamma,h}}(u_j)^2 (u_{j}-u_{j-1}) +\frac{1}{2} \sum_{j=1}^q  \oZ_{n,k,h,\ow_{\gamma,h}}(u_j)^2(u_{j+1}-u_{j}) \\
&\quad + \oZ_{n,k,h,\ow_{\gamma,h}}(u_1)^2\, \frac{u_1}{2} + \oZ_{n,k,h,\ow_{\gamma,h}}(u_q)^2 \,\frac{1-u_q}{2},
\end{align*}
where $u_0=0$ and $u_{q+1}=1$. This shows that the right hand side of \eqref{eq:aprox} is approximately equal to the mean of the left and right Riemann sum of the corresponding integral, and it is exactly this mean if $u_1=0$ and $u_q=1$. If additionally $\{u_j\}_{j=1}^q$ is a set of $q$ equidistant points in $[0,1]$, i.e. $u_j=(j-1)/(q-1)$, $j=1,\dots,q$, as in our simulation study, then the formula simplifies to
\begin{align*}
\sum_{j=1}^{q}&\oZ_{n,k,h,\ow_{\gamma,h}}(u_j)^2  (v_{j}-v_{j-1})\\
&= \frac{1}{q-1}\left[\frac{1}{2}\oZ_{n,k,h,\ow_{\gamma,h}}(u_1)^2+\sum_{j=2}^{q-1} \oZ_{n,k,h,\ow_{\gamma,h}}(u_j)^2 + \frac{1}{2} \oZ_{n,k,h,\ow_{\gamma,h}}(u_q)^2\right].
\end{align*}

To sum up, we compute 
\begin{equation}\label{eq:T.hat}
\woT_{n,h,\ow_{\gamma,h}} = \max_{1\leqslant k <n} \sum_{j=1}^{q} \oZ_{n,k,h,\ow_{\gamma,h}}(u_j)^2 (v_{j}-v_{j-1})
\end{equation}
and reject the null hypothesis for large values of $\woT_{n,h,\ow_{\gamma,h}} $, where we discuss the assessment of the significance 
in detail in the next two sections.

\subsection{Plain vanilla permutation test}

Under $H_0$, the observations $X_1,\dots,X_n$ and the observation sets $\mathcal{O}_1,\dots,\mathcal{O}_n$ are both i.i.d.\ and mutually independent, so the pairs $(X_i,\mathcal{O}_i)$ are exchangeable. 
In this work, we propose to compute the $p$-value of the test using permutation principles rather than using asymptotic critical values. A permutation approach has the benefit that the corresponding test holds its level exactly also in small sample sizes for exchangeable data, while asymptotic tests often only exhibit good size behavior for larger sample sizes. In particular, in connection with only partially observed data, the latter is an even more severe restriction as the effective sample size $N(u_j)$ of observed samples at any point $u_j$ may be significantly smaller than the total sample size $n$.

For a given statistic $T=T(Y_1,\ldots,Y_n)$ for exchangeable random variables $Y_1,\ldots,Y_n$ such as e.g.\ $\woT_{n,h,\ow_{\gamma,h}}$ for $Y_j=(X_j,\mathcal{O}_j)$, $j=1,\ldots,N$, the permutation $p$-value is obtained as
\begin{align}\label{eq:theoretic:perm}p=\Prob^*\big(T(Y_{\Pi_1},\ldots,Y_{\Pi_n})>T(Y_1,\ldots,Y_n)\big),\qquad \Prob^*(\cdot)=\Prob(\cdot|\,Y_1,\ldots,Y_n),
\end{align}
where $\boldsymbol{\Pi}=(\Pi_1,\ldots,\Pi_n)$ is a random permutation uniformly distributed over the set $S_n$ of all permutations of $(1,\ldots,n)$, i.e.\ with $\Prob(\boldsymbol{\Pi}=\boldsymbol{\pi})=1/n!$ for any $\boldsymbol{\pi}\in S_n$, independent of $Y_1,\ldots,Y_n$. A test rejecting for $p<\alpha$ is guaranteed to hold the level $\alpha$  for any sample size. 
Theoretically, conditional probabilities as required for the calculation of the permutation $p$-value as in \eqref{eq:theoretic:perm} can be calculated exactly for arbitrary $\xi$ by
\begin{align*}
    \Prob^*(T(Y_{\Pi_1},\ldots,Y_{\Pi_n})>\xi)=\frac{1}{n!}\sum_{\tpi\in\Pi_n}\ind[T(Y_{\pi_1},\ldots,Y_{\pi_n})>\xi].
\end{align*}
However, in practice, it is typically too computationally expensive to evaluate all $n!$ summands. Therefore,  one usually uses plain-vanilla Monte-Carlo approximations instead, which are based  on $B$ i.i.d.\ random permutations $\tpi^{(1)},\ldots,\tpi^{(B)}$,   uniformly distributed over $S_n$ and independent of the original data, for a fixed number $B$. 
The approximate permutation $p$-value, based on the plain-vanilla Monte-Carlo approach, for a given realization $y_1,\ldots,y_n$ of $Y_1,\ldots,Y_n$ is then obtained as
\begin{align*}
    \widehat{p}_B=\frac{1}{B}\sum_{j=1}^B\ind\left[T\big(y_{\pi^{(j)}_1},\ldots,y_{\pi^{(j)}_n}\big)>T(y_1,\ldots,y_n)\right],
\end{align*}
where $\tpi^{(j)} = (\pi^{(j)}_1,\ldots,\pi_n^{(j)})$, $j=1,\dots,B$.

Clearly, the accuracy of the permutation $p$-value $\widehat{p}_B$ depends on the number of permutation samples, so $B$ needs to be chosen quite large, usually at least $1\,000$ often even $10\,000$, such that this approach is computationally still rather costly. 
Furthermore, even if $B$ is large, it can happen, with a non-negligible probability, that $\widehat{p}_B>\alpha$, even though $p<\alpha$, or vice versa, in particular if the true $p$-value is close to the decision boundary $\alpha$, see \cite{gandy2009} for a simple illustration of this phenomenon. This is a reason for considering a modified permutation approach described in the next section.

\subsection{Sequential Monte-Carlo permutation test with bounded resampling risk}\label{sec:perm2}

We will use the same notation as in the previous section. Let $\alpha\in(0,1)$ be the chosen test level, $t=T(y_1,\ldots,y_n)$ and 
define  the partial sum process
\begin{align}\label{eq_partial_sum_perm}
S_\ell=\sum_{b=1}^\ell \ind[T(y_{\pi_1},\ldots,y_{\pi_n})>t].
\end{align}
As discussed in the previous section, the classical Monte-Carlo implementation of a permutation test is based on the approximated permutation $p$-value $\hat{p}_B=S_B/B$, where $B$ is fixed and pre-specified by the user. An alternative approach has been proposed by
\cite{gandy2009}, who shows that it is more beneficial and efficient to proceed sequentially and base the final decision rather on $S_{\tau}$, where $\tau$ is a suitably chosen random stopping time. 
More precisely, for a test decision at level $\alpha$, the approach constructs a lower and upper boundary sequence, $L_{\ell}$ and $U_{\ell}$, $\ell\ge 1$, respectively, and stops at time $\tau=\tau_{\alpha}$ as soon as the sequence $\{S_\ell\}_{\ell=1}^{\infty}$ hits one of these boundary sequences. If it hits the lower boundary, one can be sufficiently certain, that the true permutation $p$-value is indeed smaller than $\alpha$ such that the test rejects the null hypothesis. On the other hand, if it hits the upper boundary, then one can be sufficiently certain, that the true permutation $p$-value is larger than $\alpha$ such that the test does not reject the null hypothesis. The construction of these boundary curves is equivalent to a  sequential test with level $\epsilon$ 
of i.i.d.\ Bernoulli random variables with the null hypothesis that the unknown success probability equals $\alpha$.  This construction guarantees that the partial sum process $S_{\tau}$ will be on the correct side of the decision boundary $\alpha$ at the time of stopping with a probability larger than $1-\epsilon$. The value of $\epsilon$ is to be chosen by the user in advance and will be called  the tolerance. The tolerance gives the maximal probability of reaching  a different test decision than the theoretic level $\alpha$ permutation test based on the $p$-value in \eqref{eq:theoretic:perm}. Note, that this has nothing to do with the type-I- or type-II errors of the initial testing problem, but only controls the additional uncertainty due the Monte-Carlo procedure required to approximate the theoretic permutation $p$-value.
The further the  true $p$-value from \eqref{eq:theoretic:perm} is from the decision boundary $\alpha$, the quicker the procedure will typically stop. Often a decision can be reached substantially earlier than for standard choices of $B$ in the plain-vanilla Monte-Carlo procedure. On the other hand, if the true $p$-value is close to the decision boundary $\alpha$, it can take quite long until the procedure stops. Indeed, these are exactly the situations, where the plain-vanilla approach has a substantial probability of reaching a different test decision than the theoretic permutation test. In the extreme case, where the true $p$-value is exactly equal to the decision boundary $\alpha$, the procedure will never stop with a probability of at least $1-\epsilon$ for the tolerance $\epsilon$. This holds irrespective of how the boundaries are constructed as it is equivalent to the underlying sequential test for i.i.d.\ Bernoullis having a maximal type-I-error of $\epsilon$.

Essentially, the above procedure either outputs the interval $[0,\alpha)$ or $(\alpha,1]$ containing the true $p$-value with a probability of at least $1-\epsilon$. However, in some applications, a finer conclusion is desirable, such as e.g.\ distinguishing between the intervals $[0,0.001),(0.001,0.01]$, and $(0.01,0.05]$.

 Both deficiencies are overcome by \cite{gandy2020}, where the procedure is guaranteed to have a finite stopping time for all true $p$-values and outputs a finer selection of intervals:
Generally, let $\mathcal{I}$ be a set of user-specified  subintervals  of $[0,1]$ of positive lengths, open in $[0,1]$,  that cover $[0,1]$, i.e.\ the subintervals are overlapping. The elements of $\mathcal{I}$ are referred to as p-value buckets. 
\cite{gandy2020}  propose a sequential Monte-Carlo algorithm  with an output $J\in\mathcal{I}$  such that the resampling risk, i.e. the probability that the true p-value $p$ lies outside $J$, is bounded by a user-specified tolerance $\epsilon\in(0,0.5)$ for all $p\in (0,1)$. The authors further show that the expected run time  is finite if the buckets are overlapping. Hence, instead of taking only two buckets $[0,0.05)$ and $(0.05,1]$ one needs to consider, for instance, 
\begin{equation}\label{eq:I}
\mathcal{I} = \big\{ [0,0.05),(0.04,0.06), (0.05,1]\big\},
\end{equation}
where the first bucket can be interpreted as ``significant'', the last as ``non-significant'', and the middle one as ``inconclusive'' or ``borderline''.  This is the choice we use in our simulation study. However, it is also possible to add further intervals to approximate the star-classification, that is common in statistical software, such as using $[0,10^{-3}),(10^{-3},0.01),(0.01,0.05),(0.05,0.1),(0.1,1)$ in addition to some smaller ones centered at  the split points $10^{-3}$, $0.01$, $0.05$ and $0.1$,  as suggested by \cite{gandy2020}, see also the application in Section~\ref{sec:data}.

The construction of the procedure is similar to above with the difference that boundary curves are now needed for all split point between buckets, i.e.\ for $0.04$, $0.05$ and $0.06$ for $\mathcal{I}$ as in \eqref{eq:I}. Their construction is again 
associated to sequential tests (at level $\epsilon$) for the null hypothesis that the unknown success probability of the Bernoulli distribution equals the given split point. Essentially, starting with $C=(0,1)$ the algorithm proceeds as above. Whenever a boundary is hit, the set $C$ becomes smaller, e.g.\ if  $C= (s_-,s_+)$ and the lower boundary of some $s_*$ with $s_-<s_*<s_+$ is hit, then we set $C=(s_-,s_*)$. The procedure stops as soon as $C$ is contained in one of the buckets, outputting that bucket.
 It is important that the underlying sequential testing procedure fulfills a certain monotonicity assumption stating that the upper (lower) boundary curve constructed for a split points $s_1$ is always below the upper (lower) boundary curve constructed for a larger split point $s_2>s_1$. This guarantees, that the probability of outputting a bucket that does not contain the true $p$-value is upper bounded by the tolerance $\epsilon$, despite the underlying multiple testing problem for the sequential tests.
For more details, we refer to \cite{gandy2020} or to \citet[Section 4.1]{gnettner2025semcd} for an application of the same ideas in a depth-context.

\section{Simulations}\label{sec:simul}

The small sample performance of the proposed functional change point test for partially observed data is explored for $D=[0,1]$ via a Monte Carlo simulation study.
All presented results consider the observation grid $\{u_j\}_{j=1}^q$ being a set of $q=100$ equidistant points in $[0,1]$, i.e. $u_j=(j-1)/99$ for $j=1,\dots,100$.
The following variants of the test statistic $\woT_{n,h,\ow_{\gamma,h}}$  from \eqref{eq:T.hat} are considered:

\begin{itemize}
    \item[$-$] $\hoTa$, the test statistics designed for an abrupt change based on $\oZka$ in \eqref{eq:Z:abr},
    \item[$-$] $\hoTlinS$,  the test statistics designed for a linear gradual change with sum-type weights based on $\oZklinS$,
     \item[$-$] $\hoTlinI$,  the test statistics designed for a linear gradual change with the appro\-ximating integral-type weights based on $\oZklinI$.
\end{itemize}
The test statistic $\hoTlinI$, based on the weights obtained via the integral approximation, is considered, because its computation is less time consuming (roughly half the time on a personal laptop) compared to $\hoTlinS$ with the sum-type weights \eqref{eq:w-grad1}. Indeed, unlike the former, the latter does not only depend on $h_{\operatorname{lin}}$  but also in a complicated way on the missingness $\{O_i(u):i=1,\ldots,n\}$ at each location $u$. In the abrupt change case, however, the use of the approximate weights does not yield any computational gain  due to the simplicity of $w_{\gamma,\operatorname{abr}}$ and $\ow_{\gamma,\operatorname{abr}}$ in \eqref{eq_weight}, which is why we only consider the sum-type weights $\ow_{\gamma,\operatorname{abr}}$.

The performance of $\hoTa$, $\hoTlinS$ and $\hoTlinI$, is compared for the tuning para\-meters $\gamma\in\{0,1/4,1/2\}$. For $\gamma=0$  there is no weighting, so $\hoTlinSg{0}=\hoTlinIg{0}$.

 The permutation test is conducted for the overlapping buckets as in \eqref{eq:I}
with a tolerance parameter of $\epsilon=10^{-3}$  and boundaries as proposed in \cite{gandy2009} in combination with the Simctest approach, see \citet[Section 3.2]{gandy2020}, using the implementation in the \textsf{R}-package \textsf{simctest} \citep{simctest}. The empirical level and empirical power of the test are  computed for $\alpha=0.05$ from $N=1\,000$ replications as the proportion of runs with the output bucket $[0,0.05)$. 

 The setup for generation of the data is described in Section~\ref{sec:simul:1}, while  the obtained results are presented in Sections~\ref{sec:simul:2}--\ref{sec:simul:G}.  

\subsection{Data generation setup}\label{sec:simul:1}

The  variables $\{\eta_i\}_{i=1}^n$ are generated independently as
\[
\eta_i(u) = \sum_{j=0}^{J} \sqrt{\lambda_j}\, \xi_{i,j} \cos(j \pi u), \quad i=1,\dots,n,
\]
for $J=20$, $\lambda_i = 0.5 \cdot 3^{-i}$, and 
  $\xi_{i,0},\dots,\xi_{i,J}$ i.i.d.\ standard normally distributed random variables.  
Under the null hypothesis,
$X_i = \eta_i$, while under the alternative, 
\[
X_i(u)= \delta(u)\, g_u(i/n)+ \eta_i(u), \quad u\in[0,1],
\]
where $\delta$ is a function such that $\|\delta\|>0$ 
and $g_u$ is either
\begin{itemize}
\item 
$g_u(t)= h_{\operatorname{abr}}(t-\kappa) = \ind[t>\kappa]$ (abrupt change), or
\item 
$g_u(t) = h_{\operatorname{pol},r}(t-\kappa)=(t-\kappa)_+^r$ for $r\in\{1/2,1,2\}$
(gradual polynomial change). 
\end{itemize}
In both scenarios, 
$\kappa \in \{0.25, 0.50, 0.75\}$ is the change point in rescaled time. 
The sample size is chosen as $n\in\{20,50,80\}$.

\begin{figure}[tbp]
 \centering
 \includegraphics[width=\textwidth]{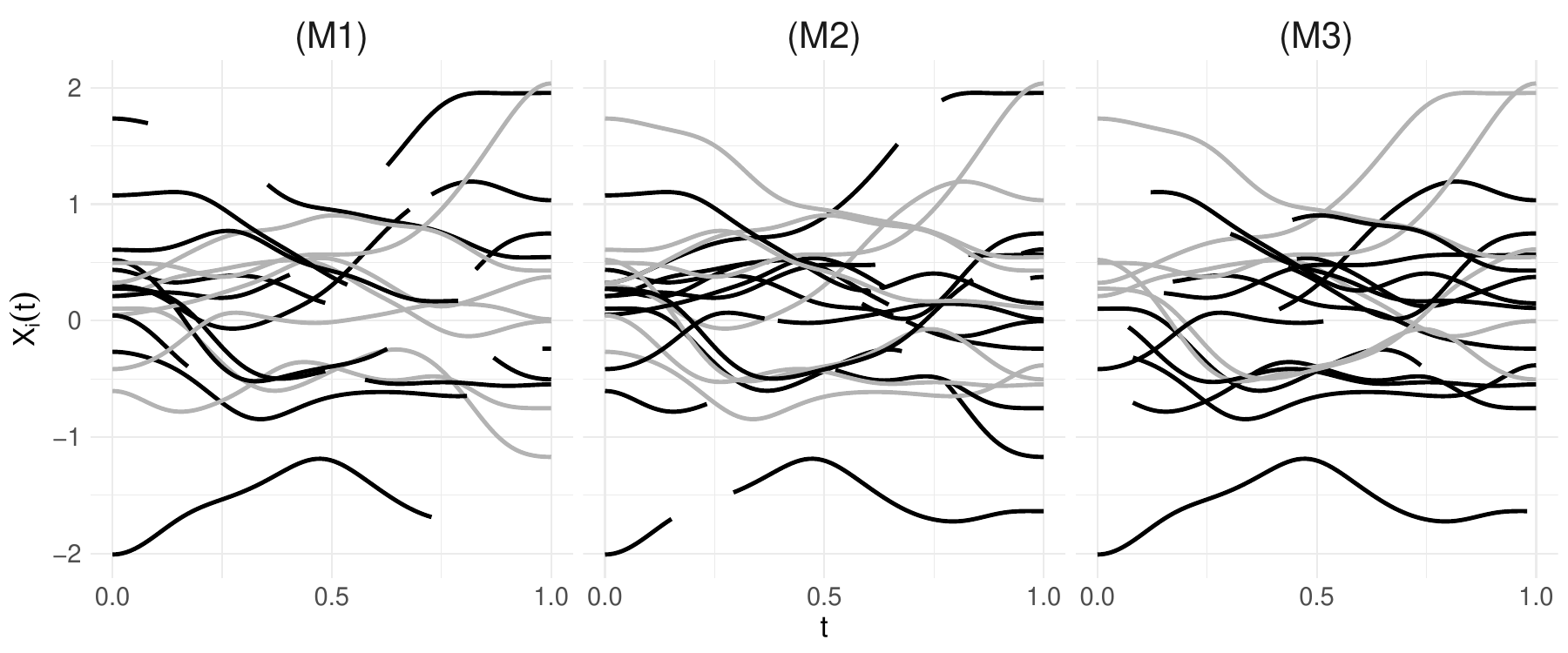}
 \caption{An example of the functional data $X_1,\dots,X_n$ with the three missingness patterns (M1)--(M3), under the null hypothesis. Each plot contains the same $n=20$ functions generated as described in Section~\ref{sec:simul:1} with observation sets (M1), (M2), and (M3), respectively. Complete profiles are in gray and incomplete ones are black.
 }\label{fig:mis2}
 \end{figure}

The observation sets $\mathcal O_i$ are
generated by one of the three missingness scenarios (M1), (M2), or (M3), that are described in detail in Section~\ref{sec:missingness} in the Supplementary file. Results for the complete data (C) are presented as a benchmark.
The missingness scenarios (M1)--(M3) are designed such that  a complete profile is observed on the grid $\{u_j\}_{j=1}^{100}$, $u_j=j/99, j=0,\dots,99$, with probability approximately 30~\% in all scenarios. 
Figure~\ref{fig:mis} in the Supplementary file shows the estimated probability of observing  $X_i(u)$ as a function of $u$ for the three scenarios.  
The models (M1), (M2) and (M3) correspond to the situations when the missingness, on average,  increases in $u$, is the strongest around $u=3/4$, and is the largest at the edges of $[0,1]$, respectively. 
Figure~\ref{fig:mis2} provides
 an illustration of the resulting functional data $\mathcal{X}_D$ with these three observation patterns.


 \subsection{Results under the null hypothesis}\label{sec:simul:2}

The empirical size of $\hoTa$, $\hoTlinS$ and $\hoTlinI$
 is presented in Table~\ref{tab:H0}. The results confirm that the test keeps the prescribed level. Results for the missing data in (M1)--(M3) seem to be comparable with the size of the test for complete data (C). 
Recall that the table only reports the percentage of runs with the output bucket $[0,0.05)$, such that it is not surprising that the empirical size is smaller than $0.05$ in most of the considered settings due to the possibility of getting the output $(0.04,0.06)$ for $p\in(0.04,0.05]$.  The relative frequency of this output
ranges from $0.004$ to $0.019$, see Table~\ref{tab:H0-add} in the Supplementary file 
for more detailed results.

\begin{table}[tb]
\centering \small
\begin{tabular}{r|rrr|rrr|rrr}
  \toprule

 \ &\multicolumn{3}{c|}{$n=20$}&\multicolumn{3}{c|}{$n=50$}&\multicolumn{3}{c}{$n=80$}\\ 
 $\gamma$ &$0$&$1/4$&$1/2$&$0$&$1/4$&$1/2$&$0$&$1/4$&$1/2$ \\
   \midrule
 \ & \multicolumn{9}{c}{Abrupt test statistic $\hoTa$}\\ 
  \midrule
(M1) & 0.050 & 0.043 & 0.051 & 0.038 & 0.030 & 0.037 & 0.033 & 0.060 & 0.049 \\ 
  (M2) & 0.050 & 0.040 & 0.040 & 0.053 & 0.043 & 0.049 & 0.045 & 0.037 & 0.034 \\ 
  (M3) & 0.042 & 0.038 & 0.044 & 0.035 & 0.040 & 0.034 & 0.042 & 0.037 & 0.049 \\ 
  (C) & 0.044 & 0.034 & 0.061 & 0.048 & 0.055 & 0.051 & 0.048 & 0.054 & 0.050 \\ 
  \midrule
    \ & \multicolumn{9}{c}{Gradual test statistic with integral-type weights $\hoTlinS$}\\
    \midrule 
(M1) & 0.048 & 0.054 & 0.043 & 0.040 & 0.045 & 0.038 & 0.051 & 0.053 & 0.045 \\ 
(M2) & 0.043 & 0.047 & 0.048 & 0.042 & 0.037 & 0.034 & 0.045 & 0.037 & 0.051 \\ 
(M3) & 0.033 & 0.037 & 0.045 & 0.047 & 0.047 & 0.056 & 0.050 & 0.053 & 0.042 \\ 
  (C) &  0.059 & 0.037 & 0.065 & 0.045 & 0.042 & 0.050 & 0.041 & 0.037 & 0.047 \\ 
\midrule
    \ & \multicolumn{9}{c}{Gradual test statistic with sum-type weights $\hoTlinI$}\\
    \midrule
(M1) & 0.048 & 0.048 & 0.044 & 0.040 & 0.048 & 0.056 & 0.051 & 0.039 & 0.048 \\ 
 (M2) &0.043 & 0.046 & 0.052 & 0.042 & 0.032 & 0.042 & 0.045 & 0.042 & 0.048 \\ 
 (M3) & 0.033 & 0.038 & 0.047 & 0.047 & 0.041 & 0.048 & 0.050 & 0.058 & 0.064 \\ 
  (C) & 0.059 & 0.038 & 0.045 & 0.045 & 0.051 & 0.054 & 0.041 & 0.043 & 0.055 \\ 
   \bottomrule
\end{tabular}
\caption{Empirical size, computed as the proportion of runs with the output bucket $[0,0.05)$,  of the permutation change point test for sample size $n\in\{20,50,80\}$ and tuning weight parameter $\gamma\in\{0,1/4,1/2\}$. The three missingness patterns (M1)--(M3) are considered together with completely observed data (C). 
}\label{tab:H0}
\end{table}

Because we use the sequential Monte-Carlo permutation test with bounded resampling risk described in Section~\ref{sec:perm2}, the number of the permutation samples $\tau$ required for the final decision of the algorithm is random. Its distribution seems to be similar for all considered $n,\gamma$ and the missingness patterns (results not presented here) and also for the three test statistics. For instance, for $n=50$, $\gamma=1/2$ and missingness (M1), and $\hoTa$, $\tau$ ranges from $10$ to $37\,953$ with median $21$, mean $472$, and upper quartile $46$. This illustrates that the usage of the permutation test based on the proposed sequential Monte-Carlo technique
is not only better with respect to the resampling risk, but it is, on average, also more efficient than the classical plain-vanilla Monte-Carlo approach which is  typically based on $B=1\,000$ or even $B=10\,000$ samples.

 \subsection{Results for an abrupt change alternative}\label{sec:simul:A}

This section focuses on results obtained for the abrupt change point statistic  $\hoTa$ for data generated under 
the abrupt change alternative. 
The power   is provided in Tables~\ref{tab:H1-1} and \ref{tab:H1-2} for  
$\delta(u) =0.7$  and  $\delta(u) = 1.2\cdot \mathrm{e}^{-2u}$, respectively.
For a fixed sample size $n$, value of the normalized change point $\kappa$, and fixed missingness pattern (a 3-cell horizontal triplet), the largest power among all choices of $\gamma$ is shown in bold. The power for fixed $n$, $\kappa$ and $\gamma$ (a 3-cell vertical triplet) is also compared among the missingness patterns and the largest one (among all missingness patterns) is underlined. 

First of all, the power behaves as expected --- it grows with the sample size $n$ and it is the largest if the change appears in the middle of the sample, i.e. $\kappa=1/2$.  Results for $\kappa=1/4$ and $\kappa=3/4$ are quite similar for fixed $n, \delta$ and  missingness pattern.  It is interesting that the power does not seem to be affected strongly by the missingness. 

\begin{table}[bp]
\centering \small
\begin{tabular}{rr|rrr|rrr|rrr}
  \toprule
&&\multicolumn{3}{c|}{$n=20$}&\multicolumn{3}{c|}{$n=50$}&\multicolumn{3}{c}{$n=80$}\\ 
 $\kappa$ & $\gamma$ &$0$&$1/4$&$1/2$&$0$&$1/4$&$1/2$&$0$&$1/4$&$1/2$ \\
  \midrule
$1/4$ & (M1) & 0.220 & \textbf{0.236} & 0.230 & 0.514 & \textbf{0.555} & 0.518 & 0.808 & \textbf{0.843} & 0.779 \\ 
   & (M2) & \textbf{0.233} & 0.225 & 0.229 & 0.583 & \textbf{0.592} & 0.559 & \textbf{0.851} & 0.849 & \underline{0.836} \\ 
   & (M3) & 0.209 & 0.211 & \textbf{0.213} & \underline{0.592} & \textbf{0.598} & 0.559 & \underline{0.851} & \textbf{0.870} & 0.832 \\ 
   & (C) & \underline{0.234} & \underline{0.240} & \underline{\textbf{0.269}} & 0.571 & \underline{\textbf{0.648}} & \underline{0.619} & 0.848 & \underline{\textbf{0.879}} & 0.829 \\ 
   \midrule
$1/2$ & (M1) & \textbf{0.396} & 0.391 & 0.307 & \textbf{0.830} & 0.788 & 0.700 & \textbf{0.962} & 0.947 & 0.909 \\ 
   & (M2) & \textbf{0.425} & 0.392 & 0.335 & \textbf{0.855} & 0.850 & 0.752 & \textbf{0.974} & 0.962 & \underline{0.943} \\ 
   & (M3) & \textbf{0.419} & \underline{0.419} & 0.331 & \textbf{0.855} & 0.848 & 0.728 & \textbf{0.971} & 0.960 & 0.941 \\ 
   & (C) & \underline{\textbf{0.449}} & 0.404 & \underline{0.351} & \underline{\textbf{0.869}} & \underline{0.852} & \underline{0.754} & \underline{\textbf{0.980}} & \underline{0.970} & 0.922 \\ 
   \midrule
$3/4$ & (M1) & 0.202 & \textbf{0.247} & 0.218 & 0.540 & \textbf{0.556} & 0.532 & 0.804 & \textbf{0.835} & 0.794 \\ 
   & (M2) & 0.212 & 0.224 & \textbf{0.259} & \underline{0.583} & \textbf{0.613} & 0.527 & 0.836 & \underline{\textbf{0.863}} & 0.825 \\ 
   & (M3) & \underline{0.227} & 0.234 & \underline{\textbf{0.260}} & 0.576 & \textbf{0.589} & 0.564 & 0.837 & \textbf{0.859} & 0.837 \\ 
   & (C) & 0.202 & \underline{\textbf{0.261}} & 0.253 & 0.566 & \underline{\textbf{0.618}} & \underline{0.575} & \underline{\textbf{0.860}} & 0.857 & \underline{0.849} \\ 
   \bottomrule
\end{tabular}
\caption{Empirical power of the permutation change point test, computed as proportion of runs with output bucket $[0,0.05)$, for an abrupt change alternative with $\delta(u) = 0.7$,
sample size $n\in\{20,50,80\}$, tuning parameter $\gamma\in\{0,1/4,1/2\}$, and change points $k=\lfloor n\kappa\rfloor$ for $\kappa\in\{1/4,1/2,3/4\}$. The 
 three missingness patterns (M1)--(M3) are considered together with completely observed data (C). The largest power among $\gamma\in\{0,1/4,1/2\}$ is shown in bold (for each $n$, $\kappa$ and the missingness separately), while the largest power among the four missingness settings is  underlined (for each $n$, $\gamma$, and $\kappa$ separately).}\label{tab:H1-1}
\end{table}

\begin{table}[htbp]
\centering \small
\begin{tabular}{rr|rrr|rrr|rrr}
  \toprule
 \ &&\multicolumn{3}{c|}{$n=20$}&\multicolumn{3}{c|}{$n=50$}&\multicolumn{3}{c}{$n=80$}\\ 
$\kappa$& $\gamma$ &$0$&$1/4$&$1/2$&$0$&$1/4$&$1/2$&$0$&$1/4$&$1/2$ \\
  \midrule
$1/4$ & (M1) & 0.165 & 0.182 & \underline{\textbf{0.212}} & 0.438 & \textbf{0.479} & \underline{0.455} & \underline{0.747} & \underline{\textbf{0.771}} & 0.704 \\ 
   & (M2) & \underline{0.181} & \underline{\textbf{0.192}} & 0.184 & \underline{0.471} & \underline{\textbf{0.511}} & 0.439 & 0.746 & \textbf{0.770} & \underline{0.758} \\ 
   & (M3) & 0.149 & \textbf{0.169} & 0.160 & 0.396 & \textbf{0.400} & 0.398 & 0.676 & \textbf{0.680} & 0.658 \\ 
   & (C) & 0.164 & \textbf{0.179} & 0.156 & \textbf{0.450} & 0.448 & 0.423 & 0.705 & \textbf{0.748} & 0.706 \\ 
   \midrule
$1/2$ & (M1) & \textbf{0.343} & \underline{0.335} & \underline{0.280} & \textbf{0.737} & 0.721 & \underline{0.644} & \textbf{0.940} & 0.920 & 0.864 \\ 
   & (M2) & \underline{\textbf{0.347}} & 0.328 & 0.260 & \underline{0.777} & \underline{\textbf{0.791}} & 0.627 & \underline{\textbf{0.956}} & \underline{0.934} & \underline{0.869} \\ 
   & (M3) & \textbf{0.298} & 0.284 & 0.251 & \textbf{0.707} & 0.673 & 0.523 & \textbf{0.917} & 0.901 & 0.818 \\ 
   & (C) & \textbf{0.325} & 0.304 & 0.224 & \textbf{0.754} & 0.727 & 0.592 & 0.927 & \textbf{0.931} & 0.847 \\ 
   \midrule
$3/4$ & (M1) & \underline{0.181} & 0.185 & \textbf{0.192} & 0.447 & \textbf{0.485} & 0.441 & 0.729 & \textbf{0.750} & 0.691 \\ 
   & (M2) & 0.169 & \underline{0.186} & \underline{\textbf{0.193}} & \underline{\textbf{0.485}} & 0.482 & \underline{0.474} & \underline{0.739} & \underline{\textbf{0.776}} & \underline{0.736} \\ 
   & (M3) & 0.142 & \textbf{0.166} & 0.142 & 0.387 & 0.425 & \textbf{0.429} & 0.643 & \textbf{0.703} & 0.630 \\ 
   & (C) & 0.166 & \textbf{0.177} & 0.171 & 0.419 & \underline{\textbf{0.490}} & 0.427 & 0.713 & \textbf{0.735} & 0.697 \\ 
   \bottomrule
\end{tabular}
\caption{Empirical power of the permutation change point test, computed as proportion of runs with output bucket $[0,0.05)$, for an abrupt change alternative with $\delta(u) = 1.2  \mathrm{e}^{-2u}$,
sample size $n\in\{20,50,80\}$, tuning parameter $\gamma\in\{0,1/4,1/2\}$, and change points $k=\lfloor n\kappa\rfloor$ for $\kappa\in\{1/4,1/2,3/4\}$. The 
 three missingness patterns (M1)--(M3) are considered together with completely observed data (C). The largest power among $\gamma\in\{0,1/4,1/2\}$ is shown in bold (for each $n$, $\kappa$ and the missingness separately), while the largest power among the four missingness settings is in red (for each $n$, $\gamma$, and $\kappa$ separately).}\label{tab:H1-2}
\end{table}

The power for different values $\gamma$ seems to be rather comparable. 
Yet, an early change with  $\kappa=1/4$ is mostly best detected  by the test statistic with $\gamma=1/4$,  with some minor exceptions. 
For the change in the middle of the sample, $\kappa=1/2$, $\gamma=0$ leads generally to the best power.
The comparison of $\gamma$'s for the late change point $\kappa=3/4$ seems similar to $\kappa=1/4$. 
Corresponding estimators for the change point and their empirical performance for the correctly specified abrupt change situation are discussed in Section~\ref{sec:estimator} in 
 the Supplementary file.

\subsection{Results for a gradual change: Weighting}\label{sec:simul:GW}

 In this section, we compare the performance of $\hoTlinS$ with the sum-type weights with the performance of the test statistic with approximate integral-type weights $\hoTlinI$ for data generated with a gradual linear change, i.e.\ the situation, where the true change function and the one used for the construction of the test statistic coincide.
 As already mentioned, the reason for considering the approximate integral-type weights is their  superior performance in terms of computational time. 
 In this section, we only consider missingness scenario (M1). The power is compared for $\kappa\in\{1/4,1/2,3/4\}$.  The function $\delta(u)$ is chosen as constant $\delta$ such that $\delta\cdot (1-\kappa)=1$, so that the strength of the signal after the change is comparable for the three choices of~$\kappa$. Since for $\gamma=0$ the two test statistics coincide, we consider solely $\gamma\in\{1/4,1/2\}$.
 The obtained results are presented in Table~\ref{tab:weight}.
 For fixed $n$, $\kappa$, and $\gamma$, the larger of the two powers is shown in bold.

\begin{table}[tbp]
\centering \small 
\begin{tabular}{rr|rrr|rrr}
  \toprule
 &\ &\multicolumn{3}{c|}{$\gamma=1/4$}&\multicolumn{3}{c}{$\gamma=1/2$}\\ 
$\kappa$ &$n$ &$20$&$50$&$80$&$20$&$50$&$80$ \\
  \midrule
$1/4$ & $\hoTlinI$ & 0.444 & 0.830 & \textbf{0.967} & 0.098 & 0.277 & 0.481  \\ 
   & $\hoTlinS$  & \textbf{0.465} & \textbf{0.845} & 0.966 & \textbf{0.344} & \textbf{0.769} & \textbf{0.936} \\ 
   \midrule
$1/2$ & $\hoTlinI$ & \textbf{0.424} & \textbf{0.795} & \textbf{0.930} & 0.121 & 0.303 & 0.493  \\ 
   & $\hoTlinS$  & 0.369 & 0.753 & 0.929 & \textbf{0.410} & \textbf{0.768} & \textbf{0.924} \\ 
\midrule
$3/4$& $\hoTlinI$ & 0.192 & 0.380 & 0.609 & 0.154 & 0.282 &0.357  \\ 
   & $\hoTlinS$  & \textbf{0.195} & \textbf{0.392} & \textbf{0.615}  & \textbf{0.295} & \textbf{0.572} & \textbf{0.792} \\ 
   \bottomrule
  \end{tabular}
  \caption{Empirical power, computed as proportion of runs with output bucket $[0,0.05)$, of test statistics $\hoTlinI$ and    $\hoTlinS$ for data generated with a gradual linear change with
  change point $k=\lfloor n \kappa \rfloor$ for $\kappa\in\{1/4,1/2,3/4\}$ with $\delta(u)=\delta$ such that $\delta(1-\kappa)=1$, and missingness pattern (M1).}\label{tab:weight}
\end{table}

The results show that the power of the two test statistics, $\hoTlinI$ and    $\hoTlinS$, is rather comparable for the weight parameter $\gamma=1/4$. The observed differences are very small, and the test statistic $\hoTlinI$ with the approximate integral-type weights appears to be even very slightly preferable in the case of a change in the middle ($\kappa=1/2$). However, the situation is completely different for the weighting with $\gamma=1/2$, where the use of approximate integral-type weights leads to a considerable loss of power, and the  statistic $\hoTlinS$ achieves, in some cases, more than twice the power of $\hoTlinI$. This may be attributed to the fact, that the equivalent statistic with $\gamma=1/2$ for univariate data with the sum-type weighting is the likelihood ratio statistic in the model where $\eta_i$ are i.i.d. standard normal, see \citet[page 59]{huskova2000}.
Comparing the results in Table~\ref{tab:weight} for fixed sample size $n$ and fixed $\kappa$ leads to a conclusion that the use of the test statistic $\hoTlinI$ with $\gamma=1/4$ seems to be a good compromise leading to a reasonable power and favorable computational properties.

 \subsection{Misspecification of the change function}\label{sec:simul:G}

\begin{table}[htb]
\centering \small
\begin{tabular}{rrr|rrr|rrr|rrr}
  \toprule
 \ &&&\multicolumn{3}{c|}{$r=1/2$}&\multicolumn{3}{c|}{$r=1$}&\multicolumn{3}{c}{$r=2$} \\ 
 $\kappa$& & $n$ &20&50&80&20&50&80&20&50&80\\
  \midrule
$1/4$ & (M1) & $\hoTag{1/4}$ & 0.407 & 0.835 & 0.962 & 0.406 & 0.792 & 0.945 & 0.307 & 0.694 & 0.859 \\ 
   &  & $\hoTlinIg{1/4}$ & \textbf{0.463} & \textbf{0.861} & \textbf{0.979} & \textbf{0.444} & \textbf{0.830} & \textbf{0.967} & \textbf{0.333} & \textbf{0.765} & \textbf{0.904} \\ 
   \cmidrule(lr){4-12}
 & (M2) & $\hoTag{1/4}$ & 0.438 & 0.851 & 0.979 & 0.412 & 0.820 & 0.968 & 0.340 & 0.702 & 0.891 \\ 
   &  & $\hoTlinIg{1/4}$ & \textbf{0.526} & \textbf{0.910} & \textbf{0.983} & \textbf{0.478} & \textbf{0.888} & \textbf{0.975} & \textbf{0.364} & \textbf{0.807} & \textbf{0.932} \\ 
    \cmidrule(lr){4-12}
 & (M3) & $\hoTag{1/4}$ & 0.434 & 0.863 & 0.976 & 0.411 & 0.834 & 0.966 & 0.311 & 0.725 & 0.887 \\ 
   &  & $\hoTlinIg{1/4}$ & \textbf{0.501} & \textbf{0.900} & \textbf{0.986} & \textbf{0.472} & \textbf{0.879} & \textbf{0.978} & \textbf{0.394} & \textbf{0.787} & \textbf{0.922} \\ 
   \cmidrule(lr){4-12}
 & (C) & $\hoTag{1/4}$ & 0.457 & 0.869 & 0.979 & 0.410 & 0.851 & 0.978 & 0.356 & 0.746 & 0.919 \\ 
   &  & $\hoTlinIg{1/4}$ & \textbf{0.514} & \textbf{0.918} & \textbf{0.989} & \textbf{0.489} & \textbf{0.898} & \textbf{0.982} & \textbf{0.401} & \textbf{0.801} & \textbf{0.948} \\ 
   \midrule
$1/2$ & (M1) & $\hoTag{1/4}$ & 0.486 & 0.863 & \textbf{0.983} & 0.350 & 0.763 & 0.926 & 0.241 & 0.529 & 0.745 \\ 
   &  & $\hoTlinIg{1/4}$ & \textbf{0.495} & \textbf{0.886} & 0.983 & \textbf{0.424} & \textbf{0.795} & \textbf{0.930} & \textbf{0.271} & \textbf{0.540} & \textbf{0.754} \\ 
   \cmidrule(lr){4-12}
 & (M2) & $\hoTag{1/4}$ & 0.490 & 0.915 & 0.986 & 0.394 & 0.780 & 0.938 & 0.242 & 0.570 & 0.799 \\ 
   &  & $\hoTlinIg{1/4}$ & \textbf{0.550} & \textbf{0.919} & \textbf{0.988} & \textbf{0.452} & \textbf{0.803} & \textbf{0.944} & \textbf{0.268} & \textbf{0.576} & \textbf{0.799} \\ 
    \cmidrule(lr){4-12}
 & (M3) & $\hoTag{1/4}$ & 0.490 & 0.899 & 0.988 & 0.376 & \textbf{0.805} & 0.951 & 0.244 & 0.544 & \textbf{0.806} \\ 
   &  & $\hoTlinIg{1/4}$ & \textbf{0.568} & \textbf{0.930} & \textbf{0.990} & \textbf{0.415} & 0.803 & \textbf{0.952} & \textbf{0.263} & \textbf{0.582} & 0.801 \\ 
   \cmidrule(lr){4-12}
 & (C) & $\hoTag{1/4}$ & 0.504 & 0.913 & 0.986 & 0.384 & 0.811 & 0.949 & 0.265 & 0.606 & \textbf{0.807} \\ 
   &  & $\hoTlinIg{1/4}$ & \textbf{0.576} & \textbf{0.922} & \textbf{0.993} & \textbf{0.460} & \textbf{0.828} & \textbf{0.961} & \textbf{0.281} & \textbf{0.636} & 0.802 \\ 
   \midrule
$3/4$ & (M1) & $\hoTag{1/4}$ & 0.264 & \textbf{0.598} & \textbf{0.850} & \textbf{0.202} & \textbf{0.381} & \textbf{0.680} & 0.107 & 0.188 & \textbf{0.376} \\ 
   &  & $\hoTlinIg{1/4}$ & \textbf{0.289} & 0.544 & 0.835 & 0.192 & 0.380 & 0.609 & \textbf{0.142} & \textbf{0.207} & 0.342 \\ 
   \cmidrule(lr){4-12}
 & (M2) & $\hoTag{1/4}$ & 0.283 & \textbf{0.644} & \textbf{0.900} & 0.213 & 0.395 & \textbf{0.701} & 0.111 & \textbf{0.236} & \textbf{0.406} \\ 
   &  & $\hoTlinIg{1/4}$ & \textbf{0.295} & 0.583 & 0.839 & \textbf{0.242} & \textbf{0.402} & 0.638 & \textbf{0.145} & 0.205 & 0.386 \\ 
    \cmidrule(lr){4-12}
 & (M3) & $\hoTag{1/4}$ & 0.284 & \textbf{0.630} & \textbf{0.870} & 0.194 & \textbf{0.404} & \textbf{0.717} & 0.121 & 0.211 & \textbf{0.440} \\ 
   &  & $\hoTlinIg{1/4}$ & \textbf{0.303} & 0.605 & 0.835 & \textbf{0.208} & 0.397 & 0.634 & \textbf{0.135} & \textbf{0.215} & 0.383 \\ 
   \cmidrule(lr){4-12}
 & (C) & $\hoTag{1/4}$ & 0.293 & \textbf{0.651} & \textbf{0.907} & 0.171 & \textbf{0.455} & \textbf{0.719} & 0.120 & \textbf{0.219} & \textbf{0.442} \\ 
   &  & $\hoTlinIg{1/4}$ & \textbf{0.330} & 0.628 & 0.858 & \textbf{0.202} & 0.408 & 0.688 & \textbf{0.139} & 0.218 & 0.370 \\ 
   \bottomrule
\end{tabular}
\caption{Empirical power, computed as proportion of runs with output bucket $[0,0.05)$, of the test statistics $\hoTag{1/4}$ and    $\hoTlinIg{1/4}$ for data generated with a gradual polynomial trend with exponent $r\in\{1/2,1,2\}$ and change point at $k=\lfloor n \kappa \rfloor$ for $\kappa\in\{1/4,1/2,3/4\}$ with
$\delta(u)=\delta$ such that $\delta(1-\kappa)^r=1$, and the four missingness patterns (M1)--(C).} \label{tab:AG}
\end{table}
\afterpage{\clearpage}

In this section, we focus on investigating the influence of misspecification of the change function in the construction of the test statistic on the power. To this end, we  further restrict to the choice $\gamma=1/4$ and compare the 
power of the abrupt change test statistic
 $\hoTag{1/4}$ with that of the test statistic designed for a linear gradual change
$\hoTlinIg{1/4}$
for data with a polynomial gradual change $h_{\operatorname{pol},r}(t-\kappa)=(t-\kappa)_+^r$ for $r\in\{1/2,1,2\}$. 
The test statistic $\hoTlinIg{1/4}$ is correctly specified if $r=1$, for all other combinations the test statistic is misspecified compared to the true change function. This is in particularly true for all combinations with $\hoTag{1/2}$, which is designed for an abrupt change.
We set  $\delta(u)=\delta$ such that $\delta (1-\kappa)^r = 1$.

The empirical power is summarized in Table~\ref{tab:AG}. For fixed $n$, $r$, $\gamma$, and $\kappa$, the larger from the two powers of $\hoTag{1/4}$ and $\hoTlinIg{1/4}$ is shown in bold. For the early change ($\kappa=1/4$) and the change in the middle of the sample ($\kappa=1/2$), the power of the gradual test statistic $\hoTlinIg{1/4}$ is typically slightly larger than the power of the abrupt test statistic $\hoTag{1/4}$, with a few exceptions where the power is almost the same. For the late change with $\kappa=3/4$,  where the sum-type approximation was also superior to the integral-type approximation, the situation differs and the comparison is less clear. Here, it appears that the gradual test statistic $\hoTlinIg{1/4}$ is beneficial for small samples ($n=20$), while the abrupt one can yield a larger power for $n\geq 50$. 

To sum up, the results indicate that the test statistic $\hoTag{1/4}$, designed for an abrupt change, leads to a reasonable power even for data with a mean that changes gradually in time instead of suddenly. In some cases (mainly for a later change), it can be even more beneficial than the test statistic designed for a gradual linear change, which could partially be attributed to the fact that we combine the gradual change statistic with the approximate integral-weights.

\section{Real data application}\label{sec:data}

The proposed methodology is illustrated using a dataset derived from a larger database on butterfly species counts in UK Environmental Change Network (ECN) terrestrial sites \citep{butterfly}.
We consider only measurements from the Drayton site (coded as T01 in the original database), which is divided into 12 sections, each with generally different vegetation and structure, and counts of 49 different butterfly species are recorded. For our analysis, we aggregate the counts of all species and sections within a single measurement.

Transect recordings were conducted weekly each year from the beginning of April to the end of September, provided that the weather conditions fulfilled the requirements of the measurement protocol (e.g., sufficient temperature). If the weather conditions did not meet the protocol, the measurement for that week is missing. While the missingness is relatively stable, it may have slightly increased over time possibly due to effects from climate change. We have therefore run additional simulations as reported in Section~\ref{sec:sensitivity} in the Supplementary file, which confirm that subtle changes in the missingness pattern alone do not blow up the size of the procedures with a slight increase for $\gamma=1/2$. Significant test results can thus safely be attributed to changes in the butterfly counts rather than climate-change induced changes in the missingness.

   During the peak of the season, when high numbers of different species occur, there are sometimes two measurements taken within a week by two different recorders. In such cases, we take the average count for that week.
As a result, there are 361 weekly measurements for 19 consecutive years (1993--2011), with up to 28 weeks per year (weeks 12--39), of counts that range from 0 to 191 with mean 26.2, see Figure~\ref{fig:data1}.  The number
of missing observations within a year ranges between 3 and 18 with mean of 9, so that the missingness in the data is non-negligible.

Observations within one year are treated as one partially observed functional variable $\{X_i(u), u\in[0,1]\}$ with $u$ being the normalized time within the year (0 for week 12 and 1 for week 39). Observations across different years can be treated as approximately independent because measurements are interrupted during the winter. The available data consist of a sequence of $n=19$ functional observations.

The test statistics $\hoTa$, 
$\hoTlinI$ and $\hoTlinS$ were considered for $\gamma\in\{0,1/4,1/2\}$, and their significance was assessed using the sequential Monte-Carlo permutation test with bounded resampling risk with  tolerance $\epsilon=0.001$. 
Even though we aim to test the null hypothesis of no change at the $0.05$ significance level, we refine the bucket $[0.00,0.05)$ to gain better insight into the significance of the test statistic. Consequently, 
the overlapping buckets are chosen as  
\begin{align}\label{eq:buckets}
\mathcal{I}=&\Big\{[0,10^{-3}),(10^{-3},0.01),(0.01,0.05),(0.05,1]\Big\}\\
&\quad \cup \Big\{(5\cdot 10^{-4},2\cdot 10^{-3}), (8\cdot 10^{-3},1.2\cdot 10^{-2}),(4.5\cdot 10^{-2},6.5\cdot 10^{-2})\Big\}.\notag
\end{align}
 The first set in \eqref{eq:buckets} relates to the classical asterix significance rating 
 used in statistical software, while the second set ensures that the buckets are overlapping and the computations finish in a finite time, see also the discussion after \eqref{eq:I}. This particular choice is recommended in \cite{gandy2020}.

\begin{table}[tbp]
\centering 
\begin{tabular}{rr|ccc}
\toprule
 & &$\gamma=0$ &$\gamma=1/4$&$\gamma=1/2$\\
\midrule
$\hoTa$& output bucket & (0.0000, 0.0010) & (0.0005, 0.0020) &  (0.0010,  0.0100)\\  
&\% & 100&99&100\\
& $\overline{\tau}$& $48\, 848$& $96\,380$& $24\, 473$\\
&($\min \tau, \max \tau)$ & (7\,753,  101\,453)&  (28\,053, \ 218\,153) & (3\,753, \ 70\,853)\\
& $\mathrm{sd}(\tau)$ & 21\,597 & 47\,611& 13\,451\\
\midrule
$\hoTlinI$&output bucket  & (0.0100,0.0500)&  (0.0100, 0.0500) &  (0.0500,    1.0000)  \\
&\% & 100 & 83 & 100\\
& $\overline{\tau}$ & 23\,171 & 142\,711 &    57\\
&($\min \tau, \max \tau)$ & (2\,753,  64\,053)&  (5\,353, 276\,253) & ( 10,    145)\\
& $\mathrm{sd}(\tau)$ & 10\,999 & 63\,852 &   27 \\
\midrule
$\hoTlinS$&output bucket  &(0.0100,   0.0500 ) & (0.0080, 0.0120) &   (0.0100  ,  0.0500)\\
&\%& 100&99&100 \\
&$\overline{\tau}$ &  21\,946& 71\,848 & 5\,177 \\
&($\min \tau, \max \tau)$ & (3\,753,  54\,153)&  (12\,153, 143\,653) & ( 1\,053,  11\,153)\\
& $\mathrm{sd}(\tau)$ &
11\,125& 23\,769& 2\,447\\
\bottomrule
\end{tabular}
\caption{Most frequent output buckets of the permutation tests with bounded resampling risk with buckets as in \eqref{eq:buckets} for the butterfly count data, together with percentages of runs (rows denoted as \%), where this output is obtained. Descriptive statistics for the number of permutation samples $\tau$  required to reach a final decision: the mean value $\overline{\tau}$, minimum and maximu, and empirical standard deviation $\mathrm{sd}(\tau)$. All quantities computed from 100 independent runs.}\label{tab:data}
\end{table}

While the theoretic $p$-value of a permutation test as in \eqref{eq:theoretic:perm} is non-random conditional on the observations, corresponding Monte-Carlo approximations are random  by construction even for fixed observations. The sequential Monte-Carlo methodology outputs a correct bucket with a large probability (referring to the random draws of the Monte-Carlo samples) of at least $1-\epsilon$. However, many of the theoretic $p$-values are contained in two buckets, such that two valid output buckets exist and the choice between those two buckets is random, where the probability of choosing one over the other depends e.g.\ on the distance of the true permutation $p$-value to the closest end point for each of the two buckets.
To illustrate this fact, we run the sequential Monte-Carlo methodology test based on the test statistics $\hoTa, \hoTlinI$, and $\hoTlinS$, for the butterfly data independently 100 times. Table~\ref{tab:data}  
presents the most frequent output bucket for each test statistic and the percentage of runs, where this output was obtained. These percentages are equal to 100 \% in most of the cases, except $\hoTag{1/4}$ and $\hoTlinSg{1/4}$, where a different bucket was obtained for one run, and  $\hoTlinIg{1/4}$, where this happened in 17 cases. The other output bucket was always the one overlapping the left endpoints of the reported most frequent output bucket.
For comparison, we present also the estimated $p$-values obtained by the plain-vanilla permutation test based on $B=10\,000$ permutation samples in Table~\ref{tab:data2}. 
However, since the number  of the permutation samples $\tau$ required by the permutation test with bounded resampling risk is often much larger than $10\,000$, see Table~\ref{tab:data}, these $p$-values should be interpreted only with caution.
The situation where the algorithm gives more than one output bucket is exactly the situation where the plain-vanilla Monte-Carlo point estimates of the $p$-values are included in both, so this is likely not due to false output buckets but due to overlapping ones.

Even though the output buckets remain stable across different permutation tests, the number of permutation samples $\tau$, required to reach the final decision, varies, and its value affects the computational time. The minimal, maximal, and mean value of $\tau$ is also presented in Table~\ref{tab:data}, together with the empirical standard deviation.

\begin{table}[tbp]
\centering 
\begin{tabular}{r|rrr}
  \toprule
 & $\gamma=0$ & $\gamma=1/4$ & $\gamma=1/2$ \\ 
  \midrule
$\hoTa$& 0.000 & 0.001 & 0.002 \\ 
$\hoTlinI$& 0.015 & 0.011 & 0.241 \\ 
$\hoTlinS$& 0.013 & 0.009 & 0.036 \\ 
   \bottomrule
\end{tabular}
\caption{Estimated $p$-values  for the butterfly counts data, based on the classical plain-vanilla permutation test based on $B=10\,000$ permutation samples.}\label{tab:data2}
\end{table}

To sum up the results, the test based on $\hoTa$  leads to the rejection of the null hypothesis of no change for all $\gamma\in\{0,1/4,1/2\}$. 
The test statistics designed for a gradual change $\hoTlinI$ and $\hoTlinS$ yield also significant results, except of $\hoTlinIg{1/2}$, which also showed a  poor power behavior in the simulation study.

 For completeness, we also report that the estimator for the change point obtained from the abrupt test statistic as discussed in Section~\ref{sec:estimator}  in the Supplementary file, which was 
 given by $\widehat{k}_{\gamma} = 10$ for 
 all choices of $\gamma$, corresponding to year 2002. Note however, that this estimator is only consistent if the data truly contains an abrupt change while a smooth transition seems more reasonable in this case.

In conclusion, the analysis based  on the new methodology proposed in this paper suggests that butterfly counts experienced a significant change over the observation period.

\afterpage{\clearpage}

\section*{Acknowledgement}
The research was supported by the Czech Science Foundation (GA\v CR) project  22-01639K (\v{S}\'{a}rka Hudecov\'{a}) as well as the German research foundation (DFG), project KI 1443/6-1 (Claudia Kirch), within the WEAVE cooperation between the DFG and GA\v CR.

\bibliographystyle{apalike}
\setlength{\bibsep}{2pt} 
{\footnotesize
\bibliography{FunLit}
}
\clearpage

\appendix

\begin{center}
{\Large \bf Supplementary file to }\\[3mm]
{\large\bf\emph{Detection of  mean changes in partially observed \\ functional data}\\[3mm]
by\\[2mm]
    \v{S}\'arka Hudecov\'a and Claudia Kirch\vspace*{0.5cm}
} 
\end{center}
\setcounter{page}{1}

In this supplement, we describe in detail the missingness scenarios for the simulation study in Section~\ref{sec:missingness}. Then, we discuss estimators for the change point in case of a correctly specified abrupt change in Section~\ref{sec:estimator}, before reporting some additional simulation results.

\section{Description of missingness scenarios}
\label{sec:missingness}

The observation sets $\mathcal O_i$ in the simulation study are 
generated by one of the following three missingness scenarios. In their definition, we use the convention that $0\cdot A = \emptyset$, $1\cdot A=A$ and $A+\emptyset = A$ for a set $A \subset D$. 
\begin{itemize}
\item[(M1)] 
$\mathcal O_i = [0,1]\smallsetminus[L_i,H_i]$
with
\[
L_i=\frac{3}{2} \sqrt{U_{1,i}} - \frac{1}{2} U_{2,i}, \quad H_i = \frac{3}{2} \sqrt{U_{1,i}}+\frac{1}{2} U_{2,i},
\]
where $U_{1,i},U_{2,i}$ are independently sampled from the uniform distribution $\mathcal{U}[0,1]$.
\item[(M2)]  $\mathcal O_i = [0,1]\smallsetminus A_i\cdot [L_i,H_i]$ with
\[
L_i=\sqrt{\frac{V_{1,i}+V_{2,i}}{2}} - \frac{1}{5} U_{2,i}, \quad H_i=\sqrt{\frac{V_{1,i}+V_{2,i}}{2}}+\frac{1}{5} U_{2,i}
\]
where $V_{1,i},V_{2,i},U_{2,i}$ are i.i.d.\ $\mathcal{U}[0,1]$-distributed, independent of $A_i$, which is Bernoulli-distributed with success probability $0.7$.
\item[(M3)]  $\mathcal O_i=A_{i,1}\cdot[0,1]+ (1-A_{i,1})A_{i,2} \cdot \left[0, H_i\right]+  (1-A_{i,1})(1-A_{i,2}) \cdot \left[L_i,1\right]$
with
\[
L_i = \frac{1-\sqrt{U_i}}{2}, \quad H_i= \frac{1+\sqrt{U_i}}{2},
\]
$A_{i,1}$, $A_{i,2}$ are independent Bernoulli variables with success probabilities
$0.3$ and $0.5$, respectively, and independent of $U_i \sim \mathcal{U}[0,1]$. By construction only one of the three sets that make up the summands is not empty.
\item[(C)] $\mathcal O_i=[0,1]$, i.e. complete data with no missingness. 
\end{itemize}
In (M1) and (M2),  if $L_i>1$, then a full profile is being observed, which is also the case for (M2) and $A_i=0$. In (M3) a full profile is observed  for $A_{i,1}=1$. 

\begin{figure}[htbp]
 \centering
 \includegraphics[width=0.5\textwidth]{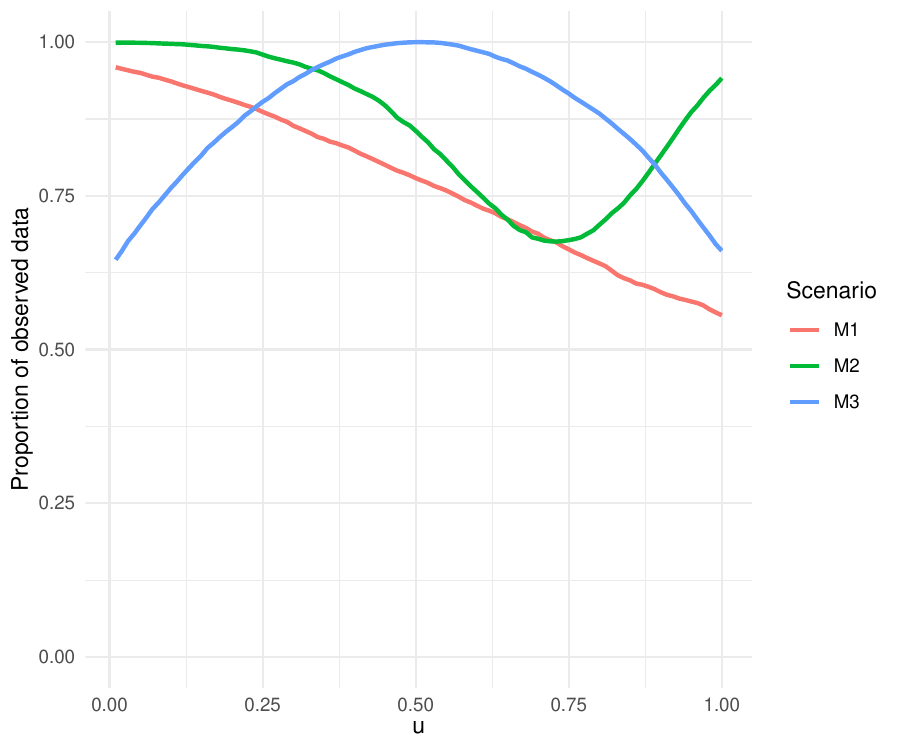}
 \caption{Estimated probability of observing $X_i(u)$ computed for each $u\in [0,1]$ for the three  scenarios (M1)--(M3) as a relative proportion in a sample of size
$n=10\,000$.}\label{fig:mis}
 \end{figure}

The missingness pattern (M1) is inspired by the simulation study from \cite{kraus2019}. 
The support of the density in (M1) allows all positions for the missing part: the beginning of $[0,1]$, its interior, its end, or no missingness. 
On the other hand,  the shape of the density in (M2) causes  the missingness to appear most likely in the interior of $[0,1]$ or at its end, and the length of the missing part is, on average,  narrower that in (M1). Moreover, 
in most of the cases $L_i<1$,
so  the proportion of completely observed curves is increased by the presence of the Bernoulli variable $A_i$ in the definition of $\mathcal{O}_i$. 
Finally, the missingness pattern (M3) is designed such that the unobserved part, if present (i.e. $A_{i,1}=0$), is either at the beginning of $[0,1]$ or at the end of $[0,1]$, with the same probability driven by $A_{i,2}$. 
The choice of the distribution for $A_i$ in (M2) and $A_{i,1}$ in (M3) is such that a complete profile is observed on a grid $\{u_j\}_{j=1}^{100}$, $u_j=j/99, j=0,\dots,99$, with probability approximately 30~\% in all scenarios. 
Figure~\ref{fig:mis} shows the estimated probability of observing  $X_i(u)$ as a function of $u$ for the three scenarios.  
The models (M1), (M2) and (M3) correspond to the situations when the missingness, on average,  increases in $u$, is the strongest around $u=3/4$, and is the largest at the edges of $[0,1]$, respectively. 

\section{Change point estimation for an abrupt change}\label{sec:estimator}

Consider the setup from Section~\ref{sec:amoc} and the corresponding test statistic 
\[
\oT_{n,\gamma,\operatorname{abr}} = \max_{1\leq k\leq n}\|\oZka\|^2
\]
for $\oZ_{n,k,\gamma,\operatorname{abr}}$ from \eqref{eq:Z:abr}.
In the correctly specified situation, i.e.\ if the true change function is indeed given by $h_{\operatorname{abr}}(t-\kappa)$ for some $0<\tau<1$,  then the change point $k=\lfloor \tau n\rfloor$ can be estimated as
\begin{equation*} 
\mathrm{argmax}_{1\leqslant k <n}  \|\oZka\|^2.
\end{equation*}
In practice, when only observations on a grid $\{u_j\}_{j=1}^q$, $u_j\in D$, $j=1,\dots,q$, are available, 
this change point estimator is calculated, when the null hypothesis is rejected, as
 \begin{equation*}
 \widehat{k}_{\gamma} =\mathrm{argmax}_{1\leqslant k <n}  \sum_{j=1}^{q} \oZka(u_j)^2 (v_{j}-v_{j-1}),
 \end{equation*}
where $v_j$, $j=0,\dots,q$, are introduced in Section~\ref{sec:4} of the main paper.

\begin{figure}[tbp]
    \centering
    \includegraphics[width=0.9\linewidth]{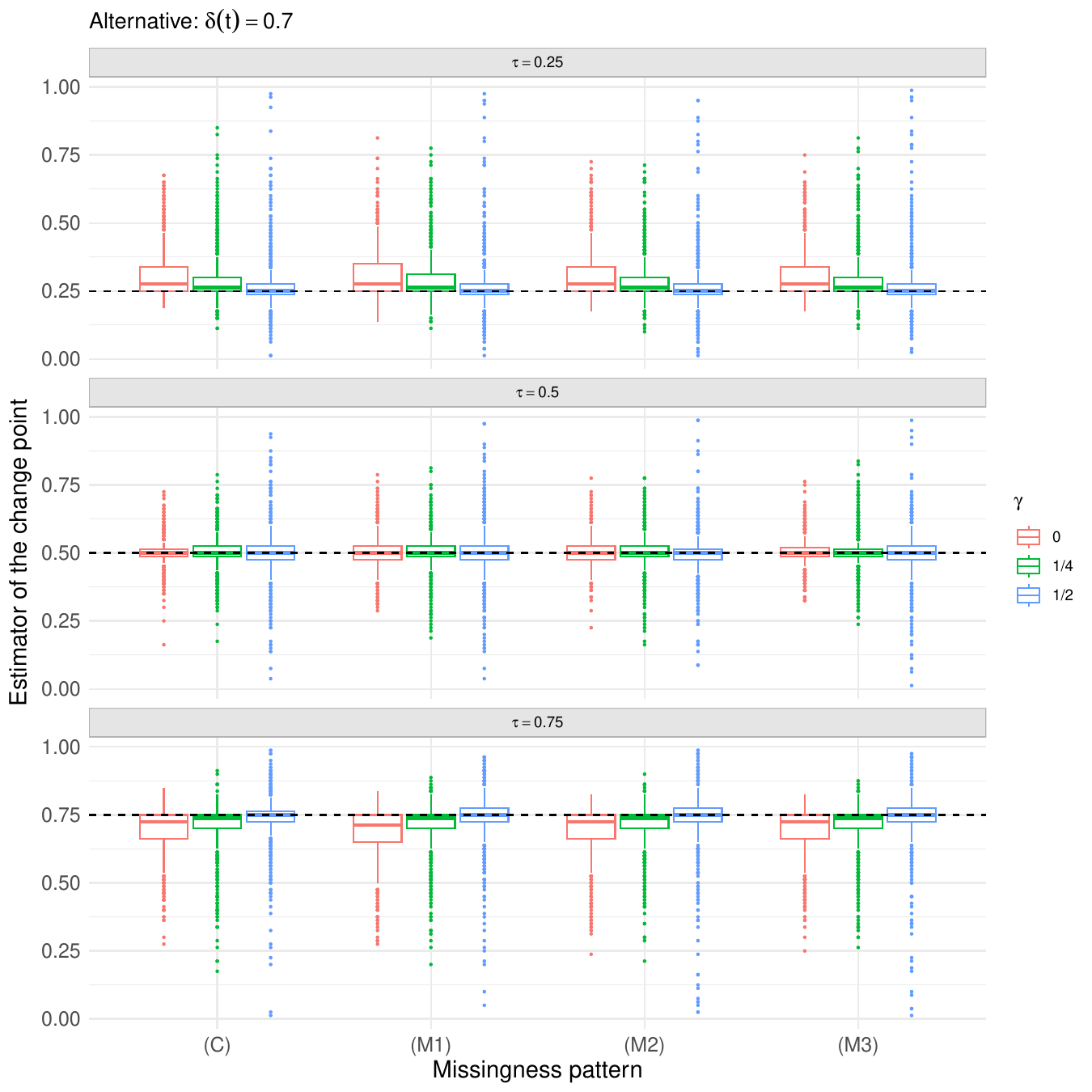}
    \caption{Boxplots of the estimator $\widehat{\kappa}_{\gamma}$,computed for significant outputs only, for the alternative $\delta(u)=0.7$, sample size $n=80$, tuning parameter $\gamma\in\{0,1/4,1/2\}$ and the three missingness patterns (M1)--(M3) and complete data (C). }
    \label{fig:1-box}
\end{figure}

The performance of the estimator $\widehat{k}_{\gamma}$ for $\gamma\in\{0,1/4,1/2\}$ and $n=80$ is examined for data generated with an abrupt change at $k=\lfloor n \kappa \rfloor$ for $\kappa\in\{1/4,1/2,3/4\}$, as in Section~\ref{sec:simul:A} of the main text. 
The distribution of  $\widehat{k}_{\gamma}$  for  $\delta(u)=0.7$ is presented  using boxplots in Figure~\ref{fig:1-box}. 
All characteristics are computed solely from the significant cases where the null hypothesis of no change is rejected.
The results indicate that the change point is generally estimated well and differences among the missingness patterns are not visible from the plots. 
The best results for $\widehat{\kappa}_{\gamma}$ are obtained if the change appears in the middle of the sample, i.e. $\kappa=1/2$. For this case, the results for different $\gamma$ values are very similar, while for $\kappa\ne 1/2$, the choice $\gamma=1/2$ seems to lead to the best performing estimator, as indicated by the boxplots in Figure~\ref{fig:1-box}. 
For an early change point with $\kappa=1/4$,  the medians for $\gamma=0$ and $\gamma=1/4$ are larger than  $\kappa$, so the method overestimates the true change point. For $\kappa=3/4$ we observe the opposite, i.e. underestimation. 
The results for 
$\delta(u)=1.2\mathrm{e}^{-2u}$ are very similar, so they are not presented here. 

\clearpage

\section{Additional results under the null hypothesis}

The empirical size of the test, as presented in Table~\ref{tab:H0} of the main text, is computed as the percentage of runs with the output bucket $[0,0.05)$. In order to complement these results, we present in Table~\ref{tab:H0-add} the relative frequencies of the bucket $(0.04,0.06)$.

\begin{table}[hb]
\centering
\begin{tabular}{l|rrr|rrr|rrr}
 \toprule
 \ &\multicolumn{3}{c|}{$n=20$}&\multicolumn{3}{c|}{$n=50$}&\multicolumn{3}{c}{$n=80$}\\ 
 $\gamma$ &$0$&$1/4$&$1/2$&$0$&$1/4$&$1/2$&$0$&$1/4$&$1/2$ \\
  \midrule
    \ & \multicolumn{9}{c}{Abrupt test statistic $\hoTa$}\\
  \midrule
(M1) & 0.004 & 0.014 & 0.011 & 0.009 & 0.018 & 0.009 & 0.011 & 0.014 & 0.007 \\ 
  (M2) & 0.011 & 0.007 & 0.019 & 0.009 & 0.014 & 0.018 & 0.006 & 0.012 & 0.012 \\ 
  (M3) & 0.010 & 0.018 & 0.010 & 0.008 & 0.012 & 0.007 & 0.014 & 0.007 & 0.012 \\ 
  (C) & 0.009 & 0.005 & 0.005 & 0.013 & 0.009 & 0.009 & 0.014 & 0.010 & 0.007 \\ 
\midrule
 \ & \multicolumn{9}{c}{Gradual test statistic with sum-type weights $\hoTlinS$}\\
  \midrule
(M1) &  0.013 & 0.014 & 0.039 & 0.011 & 0.011 & 0.004 & 0.010 & 0.012 & 0.015 \\ 
  (M2) &0.012 & 0.014 & 0.028 & 0.010 & 0.010 & 0.005 & 0.011 & 0.012 & 0.012 \\ 
  (M3) & 0.009 & 0.010 & 0.039 & 0.007 & 0.012 & 0.004 & 0.006 & 0.008 & 0.006 \\ 
  (C) & 0.007 & 0.003 & 0.018 & 0.011 & 0.009 & 0.011 & 0.016 & 0.011 & 0.003 \\ 
    \midrule
    \ & \multicolumn{9}{c}{Gradual test statistic with integral-type weights $\hoTlinI$}\\
  \midrule
(M1) & 0.013 & 0.009 & 0.013 & 0.011 & 0.011 & 0.007 & 0.010 & 0.008 & 0.008 \\ 
(M2) & 0.012 & 0.015 & 0.007 & 0.010 & 0.007 & 0.018 & 0.011 & 0.013 & 0.011 \\ 
(M3) & 0.009 & 0.004 & 0.011 & 0.007 & 0.010 & 0.011 & 0.006 & 0.017 & 0.014 \\ 
  (C) &  0.007 & 0.006 & 0.009 & 0.011 & 0.018 & 0.016 & 0.016 & 0.008 & 0.010 \\ 
   \bottomrule
\end{tabular}
\caption{Relative frequency of the output $(0.04,0.06)$ under the null hypothesis.}\label{tab:H0-add} 
\end{table}

\clearpage
\section{Sensitivity with respect to non-stable missingness}\label{sec:sensitivity}

In order to investigate the sensitivity of the testing procedure with respect to non-stable missingness, we computed the empirical size for data generated under $H_0$ and changing missingness patterns. Namely, the sets $\mathcal{O}_i$ were simulated as in (M2), but with $A_i$ being Bernoulli with success probability $p_i$, where $p_i = 0.7$ for $1\leqslant i\leqslant  n/2$ and $p_i = 0.9$ for $n/2+1 \leqslant i \leqslant n$. The results, summarized in Table~\ref{tab:H0-sens}, reveal that the proposed testing procedure is quite robust with respect to small changes in the missingness pattern.

\begin{table}[ht]
\centering \small 
\begin{tabular}{r|rrr|rrr|rrr}
  \toprule
 \ &\multicolumn{3}{c|}{$n=20$}&\multicolumn{3}{c|}{$n=50$}&\multicolumn{3}{c}{$n=80$}\\ 
 $\gamma$ &$0$&$1/4$&$1/2$&$0$&$1/4$&$1/2$&$0$&$1/4$&$1/2$ \\
  \midrule
$\hoTa$ & 0.045 & 0.049 & 0.037 & 0.048 & 0.044 & 0.042 & 0.032 & 0.046 & 0.055 \\ 
  $\hoTlinS$ & 0.046 & 0.041 & 0.049 & 0.047 & 0.050 & 0.055 & 0.050 & 0.035 & 0.041 \\ 
  $\hoTlinI$ & 0.046 & 0.055 & 0.036 & 0.047 & 0.043 & 0.064 & 0.050 & 0.043 & 0.056 \\ 
   \bottomrule
\end{tabular}
\caption{Empirical size, computed as the proportion of runs with the output bucket $[0,0.5)$, of the permutation change point test for sample size $n\in\{20,50,80\}$ and tuning weight parameter $\gamma\in\{0,1/4,1/2\}$ for data generated under $H_0$ and non-stable missingness as in (M2), but with $A_i$ being Bernoulli($p_i$), where $p_i$ changes from 0.7 to 0.9 at $k=n/2$.}\label{tab:H0-sens}
\end{table}

\end{document}